\documentclass[iop,revtex4]{emulateapj}
\usepackage[english]{babel}
\usepackage[colorinlistoftodos]{todonotes}
\usepackage{graphics,graphicx}
\usepackage{helvet}
\usepackage[utf8]{inputenc}
\usepackage[T1]{fontenc}
\usepackage{longtable}
\usepackage{soul}
\usepackage{hyperref}
\usepackage{amsmath, amssymb, gensymb}
\usepackage{ulem}
\usepackage{natbib}
\usepackage{microtype}
\usepackage{multirow}
\usepackage{morefloats}
\usepackage{rotating}
\usepackage{pdflscape}
\usepackage{color}
\usepackage{xspace}
\def\Msol {$\hbox{M}_\odot$\xspace}

\def\kms {km\,s$^{-1}$\xspace}

\def\etal {\textit{et al.}\xspace}
\def\aap{A\&A}                   % "Astron. Astrophys."
\def\apjs{ApJS}                  % "Astrophys. J. Suppl. Ser."
\def\apjl{ApJ}                   % letter at ApJ

\begin{document}

\slugcomment{\textit{Accepted for publication in ApJ on 25 Nov 2016}}

\title{Synthetic Observations of Magnetic Fields in Protostellar Cores}
\shorttitle{Synthetic Observations of Magnetic Fields in Protostellar Cores}

\author{Joyce~W.~Y.~Lee\altaffilmark{1,2}}
\author{Charles~L.~H.~Hull\altaffilmark{1,3}} 
\author{Stella~S.~R.~Offner\altaffilmark{4}}

\shortauthors{Lee \etal}
\altaffiltext{1}{Harvard-Smithsonian Center for Astrophysics, 60 Garden St., Cambridge, MA 02138, USA}
\altaffiltext{2}{Department of Physics and Astronomy, University of Southampton, University Road, Southampton, SO17 1BJ, UK; \href{mailto:jwyl1g12@soton.ac.uk}{\texttt{jwyl1g12@soton.ac.uk}}}
\altaffiltext{3}{Jansky Fellow of the National Radio Astronomy Observatory; \href{mailto:chat.hull@cfa.harvard.edu}{\texttt{chat.hull@cfa.harvard.edu}}}
\altaffiltext{4}{Department of Astronomy, University of Massachusetts, Amherst, MA 01003, USA}
\begin{abstract}
\noindent
The role of magnetic fields in the early stages of star formation is not well constrained. In order to discriminate between different star formation models, we analyze 3D magnetohydrodynamic simulations of low-mass cores and explore the correlation between magnetic field orientation and outflow orientation over time. 
We produce synthetic observations of dust polarization at resolutions comparable to millimeter-wave dust polarization maps observed by CARMA and compare these with 2D visualizations of projected magnetic field and column density.  
Cumulative distribution functions of the projected angle between the magnetic field and outflow show different degrees of alignment in simulations with differing mass-to-flux ratios. The distribution function for the less magnetized core agrees with observations finding random alignment between outflow and field orientations, while the more magnetized core exhibits stronger alignment. We find that fractional polarization increases when the system is viewed such that the magnetic field is close to the plane of the sky, and the values of fractional polarization are consistent with observational measurements. The simulation outflow, which reflects the underlying angular momentum of the accreted gas, changes direction significantly 
over the first $\sim0.1$ Myr of evolution. This movement could lead to the observed random alignment between outflows and the magnetic fields in protostellar cores. 
\\
\end{abstract}

\section{Introduction}
\label{sec:intro}

Magnetic fields have long been thought to be significant in regulating star formation \citep{Shu1987,Mckee1993}, but their influence during the gravitational collapse of a dense core remains unclear.  The collapse of clouds to form stars is regulated by a combination of magnetic and turbulent support, which act against gravity. The significance of magnetic forces is quantified by the mass-to-flux ratio, M$_\Phi $ = M/$\Phi$, or the ratio of the gravitational potential energy to magnetic energy ($B^2$/8$\pi$). A cloud must be super-critical (M$_\Phi$ > 1), i.e., the gravitational energy must exceed the magnetic energy, for a protostar to form \citep{Crutcher2012}.  A sub-critical cloud (M$_\Phi$ < 1) is magnetically supported and therefore will not collapse.e.g.,

A variety of prior observations have informed our current understanding of magnetic fields in star formation. Polarimetric observations provide evidence of magnetic fields in molecular clouds and nebulae, as oblong dust grains tend to be aligned by magnetic fields \citep[e.g.,][]{Lazarian2007}.  Observations of background starlight polarization, caused by absorption by aligned dust grains, include studies of the Musca Dark Cloud \citep{Pereyra2004}, the Pipe Nebula \citep{Franco2010} and the Lupus I molecular cloud \citep{Franco2015}.  Studies in the (sub)millimeter regime have analyzed polarized thermal emission from aligned dust grains, and span spatial scales from molecular clouds \citep{Planck2015a,Planck2015b,Planck2015c} to protostellar cores \citep{Matthews2009, Dotson2010}, envelopes \citep{Girart2006, Girart2009, Stephens2013, Hull2013, Hull2014, Zhang2014}, and protostellar disks \citep{Stephens2014, Rao2014}.  

Investigations into magnetic field orientations across spatial scales provide additional insight \citep[e.g.,][]{HuaBaiLi2009,2014A&A...569L...1A}. \citet{HuaBaiLi2009} used optical background-starlight polarimetry to probe field morphology on cloud scales (200\,pc) and sub-millimeter dust polarimetry to probe fields on much smaller core scales (a few $\times$ 0.1\,pc). They showed that magnetic-field orientations are broadly consistent from cloud- to core-scales.

A second, related open problem in star formation concerns the evolution of angular momentum from large to small spatial scales \citep{mckee07}.   Magnetic torques can provide an efficient braking mechanism and thus remove angular momentum on small scales \citep[e.g.,][]{mestel85}.  At intermediate scales, star-forming cores are observed to have rotation \citep{Goodman93, Chen2007, Tobin2011}, which may be influenced by the geometry of the local magnetic field \citep{Dib10}. However, it is not clear whether the angular momentum vector of the dense gas typically aligns with the magnetic field direction on < 1\,pc scales. On sub-AU scales the magnetic field is essential to launching and collimating outflows, although the exact mechanisms are disputed \citep{2014prpl.conf..451F,2014prpl.conf..173L}. 

In the case of tight alignment between the rotational axis and magnetic field orientation, outflows would be aligned with the field, i.e., perpendicular to a rotationally supported disk. Where there are strong fields, this aligned scenario restricts the formation of disks, resulting in the ``magnetic braking catastrophe'' \citep[e.g.,][]{2003ApJ...599..363A}. In contrast, disk growth is promoted by misalignment between the field and rotation axis \citep{Hennebelle2009,Joos2012,Krumholz2013,Li2013} and by non-ideal MHD effects such as ambipolar diffusion and Ohmic dissipation \citep[e.g.,][]{DappBasu2010, Li2011, Machida2011, MachidaMatsumoto2011, Dapp2012, Machida2013, Tomida2013}. 

The distribution of the alignment of the rotation axis and the magnetic field direction in dense gas is as yet unknown.  Observational studies have reached different conclusions regarding the alignment of fields and outflows in cores, where the outflow is used to trace the angular momentum direction of the star and the disk. Tight alignment between the outflow and magnetic field orientations has been found in studies like that of \citet{Chapman2013}, who mapped polarization towards Class 0 protostars on $\sim$\,10,000\,AU scales. However, interferometric results from a survey by CARMA (the Combined Array for Research in Millimeter-wave Astronomy) on $\sim$\,1000\,AU scales showed that outflows and magnetic fields are not tightly aligned but may be randomly aligned \citep{Hull2013,Hull2014}. This indicates that the magnetic field behavior may change across spatial scales. 

Given the observational expense and technical difficulty of studying magnetic fields, only a handful of high resolution surveys have been performed. Therefore, in search of an explanation of these results, we turn to simulations that contain full physical information at much higher resolutions than observations. For a meaningful comparison with observations we model emission via radiative transfer, creating a ``synthetic observation'' \citep{2011IAUS..270..511G}. 

Although a variety of studies have created synthetic observations of simulations \citep[e.g.,][]{ 2008AJ....136..404O, 2011IAUS..270..511G,Offner2011}, only a few studies exist of synthetic magnetic field observations of cores modeled with full 3D magnetohydrodynamics (MHD) \citep{Matsumoto2006, Tomisaka2011, Frau2011, Kataoka2012, Li2015}. In this investigation, we perform radiative transfer on 3D MHD simulations that include both radiative and outflow feedback. To understand the alignment (or lack thereof) between outflows and magnetic fields, we measure the average magnetic field orientations in synthetic dust polarization maps and the outflow direction from visualizations, and compare the synthetic results with observational studies. 

We produce visualizations and synthetic observations of two simulations of low-mass young stellar objects (YSOs) with differing mass-to-flux ratios. By viewing the simulations at different angles and times, we generate a survey of synthetic ``sources.''  We compare the polarization fractions of the synthetic maps with polarization fractions from observations. We also produce cumulative distribution functions (CDFs) of the angle between the outflow and magnetic field direction. Finally, we examine the origin of random alignment and the evolution of outflows over time. 

We first present the details of the numerical simulations (\S\ref{sec:method1}). We then explain the post-processing and the generation of synthetic dust polarization (\S\ref{sec:method2} \& \S\ref{sec:method3}). We present the results (\S\ref{sec:results}) before discussing conclusions and future work (\S\ref{sec:conclusions}).

\section{Method}

\subsection{Numerical Simulations}
\label{sec:method1}

We analyze two ideal MHD simulations of isolated, collapsing cores run by Offner et al.~(in prep). The simulations were performed using \textsc{orion}, which
solves the partial differential equations of MHD on an adaptive mesh refinement (AMR) grid \citep{PSLi2012}.  In the ideal approximation the gas and field are treated as perfectly coupled.  The simulations follow the collapse of a dense core, the subsequent formation of a protostar, and the interaction of the protostar with its surroundings through radiative feedback and a bipolar outflow.

The initial conditions and parameters are nearly identical to those described in \citet{Offner2016}, a study that focused on initial turbulent realizations that produced long-lived binary or multiple systems.  The initial configuration consists of a uniform-density spherical core of cold gas (10\,K) surrounded by a warm (1000\,K) low-density medium. The initial density of the cold gas is 2.38~$\times$~10$^{-19}$\,g\,cm$^{-3}$. The initial core has a mass of 4\,\Msol a radius of 0.065\,pc. The initial magnetic field is uniform and vertical in the $z$-direction. At $t=0$, the gas in the core is perturbed with a turbulent random velocity field, which is normalized to satisfy the input velocity dispersion (0.46\,\kms). The initial turbulence is allowed to decay, with no further external energy injection.  

The initial base grid resolution is $64^3$, but the dense core is refined by two additional levels via a density threshold refinement criterion. As the gas collapses, additional AMR levels are inserted when the density exceeds the Jeans condition for a Jeans number of $J$ = 0.125 \citep{1997ApJ...489L.179T}. The gas is also refined on radiation energy density gradients according to $\Delta E_r/E_r > 0.15$ and by up to two levels if $\Delta \rho/\rho > 0.6$, conditions which allow the warm circumstellar region and outflow cavity to be well-resolved.  Each subsequent level increases the cell resolution by a factor of 2. 

When the Jeans condition is violated on the fifth level, a sink particle is added \citep{2004ASPC..323..401K}. The sink particle includes a sub-grid model for radiation feedback and outflow launching \citep{Offner2009, 2011ascl.soft04002C}. The outflow is launched via an X-wind model based upon \citet{1999ApJ...526L.109M}, such that $f_{\rm out}=0.2$ of the accreting material is launched in an outflow \citep[e.g.,][]{Cunningham2011,Hansen2012,OffnerArce2014,Myers2014}. Without such a model, outflows produced in magnetized simulations with resolutions coarser than 1\,AU are slow and not well collimated or absent altogether \citep{Seifried2012, Machida2013, Kataoka2012, Joos2013}.  
 
 The model defines the outflow axis to be parallel to the net angular momentum vector of the protostar, which is set by the protostellar accretion history \citep[for details see][]{Fielding15}. The outflow velocity is given by the Keplerian velocity near the protostellar radius: $v_{\rm K} =f_{\rm K} \sqrt{Gm_p/r_p}$, where $f_{\rm K} = 0.3$, $m_p$ is the instantaneous protostellar mass, and $r_p$ is based on a stellar evolution model that incorporates the internal state and accretion history of the protostar \citep{Offner2009}. The opening angle, $\theta_p$, over which the outflow momentum is distributed on the grid, is a free parameter. However, it is constrained by observed outflow properties, and we adopt $\theta_p = 0.01$, which appears typical for low-mass protostellar outflows \citep{1999ApJ...526L.109M,Cunningham2011}. \citet{OffnerArce2014} vary $\theta_p$  and $f_{\rm out}$ and find that they have a mild impact on the overall star formation efficiency and outflow characteristics. However, these parameters do not solely determine the outflow properties; the interaction between the outflow and turbulent core envelope also impacts the large-scale outflow properties \citep[see also][]{Offner2011,Joos2013}. Here, the outflow direction is set by the angular momentum of the accreted turbulent gas \citep{Offner2016}. Since  turbulence within the core is in turn regenerated by the outflow as shown by \citet{OffnerArce2014},  the outflow and turbulence are inter-dependent and evolve in tandem. 

In our study we analyze two simulations with mass-to-flux ratios of  $M_\Phi=2.5$  and $M_\Phi=1.5$, which we coin ``S2.5'' and ``S1.5.'' These correspond to initial magnetic field strengths of B = 41 $\mu$G and B = 68 $\mu $G, respectively.  Each calculation evolves until shortly after 0.5\,Myr. The protostar forms around 0.25\,Myr in S2.5 and 0.35\,Myr in S1.5.  S2.5 forms a second protostar; however, its mass remains small ( $< 0.1\,M_\odot$). At 0.38\,Myr the secondary approaches within $\sim$\,200\,AU of the primary and merges with it.

\subsection{Post-processing} 
\label{sec:method2}
We render the 3D solutions into 2D projection plots of density, magnetic field, and energy density using the Python toolkit \texttt{yt} \citep{Turk2011}. 

In order to trace the outflow direction of the system at various time steps, we produce projections of the energy density from different viewing angles. For each time step we view the outflow from the positive $x$-, $y$-, and $z$-directions. By viewing the synthetic sources from orthogonal perspectives we eliminate biased projections, mimicking the effects of observing the system isotropically. We also examine the impact of system orientation on the perceived direction of the outflow by considering small changes in viewing angle; we view the system at $\theta = 0\degree, 30\degree, 45\degree, 60\degree$, and $90\degree$, where $\theta=0\degree$ corresponds to the view of the $x$-$y$ plane from the direction of the $z$-axis (``face-on'') and $\theta = 90\degree$ corresponds to the view of the $x$-$z$ plane from the direction of the $y$-axis (``edge-on'').  (See Figures \ref{fig:Energy density xyz} and \ref{fig:Energy density 304560}.)

Thermal dust polarization probes the magnetic field in the plane of the sky; therefore, we first create visualizations of the raw magnetic field data by calculating the 2D magnetic field direction in the plane of the sky from the 3-component magnetic field. The magnetic flux at each point in the grid is calculated by summing along the line of sight and dividing by the data resolution. This, combined with visualizations of the column density, provides information on the magnetic field orientation and total dust continuum (Stokes $I$) emission before radiative transfer. By combining a series of visualizations over a period of time, we observe the changes in magnetic field morphology over the course of the protostar's lifetime.

\subsection{Synthetic Dust Polarization (Radiative Transfer)}
\label{sec:method3}
We process the simulation data using the ray-tracing radiative transfer code \texttt{DustPol} \citep{Padovani2012}, which is part of the ``Adaptive Radiative Transfer Innovations for Submillimeter Telescopes'' (\texttt{ARTIST}) package. \texttt{ARTIST} is based on \texttt{LIME} (Line Modeling Engine) \citep{Brinch2010}, a non-local thermodynamic equilibrium (LTE) Monte Carlo radiative transfer code that allows adaptive 3D ray-tracing. \texttt{LIME} reads in values of $x$, $y,$ and $z$ in a Cartesian co-ordinate system and requires density and 3-component magnetic field information for every point in the cube. 

The ray-tracing algorithm generates a Delaunay grid that allows adaptive smoothing. The grid is defined by the number of grid points and sink points. The grid points are distributed inside the model, which returns smoother results with a higher number of grid points. The number of sink points defines the chosen points on the surface, i.e., the surface density. More time is needed for a larger  number of points; however, under-sampling creates artifacts in the image. We tested grids with different numbers of sink points and grid points to check for artifacts, converging on a Delaunay grid of 10,000 grid points and 6000 sink points.

We generate images in a range of sizes, including scales comparable to those typically viewed with CARMA ($\approx$\,1000\,AU). For our dust emission models, we set an object distance of 140 pc---approximately the same distance as the Ophiuchus molecular cloud. 

The input tables of simulation data have physical spatial resolutions that are comparable to those of CARMA observations of objects at the given distance. We use a 512 $\times$ 512 grid of pixels with a pixel resolution of 0.7$\arcsec$, smoothed to an image resolution of 3$\arcsec$, which is comparable to the 2.5$\arcsec$ beam size of the observations. We use a frequency of 245\,GHz, which is near the center of the 210--270\,GHz bandwidth of the CARMA polarization system \citep{Hull2015}.

The radiative transfer model assumes a fixed gas-to-dust ratio of 1:100; we specify dust opacity values from tables calculated using the coagulation models detailed in \citet{1994A&A...291..943O} for thin ice-mantles and gas densities on the order of $\sim$\,10$^{5}$\,cm$^{-3}$.

We use a maximum grain alignment efficiency of 15\% for gas densities with H$_2$ number density $n\,<\,10^{12}$\,cm$^{-3}$ \citep{Padovani2012}. In dense regions with $n\,>\,10^{12}$\,cm$^{-3}$ the model assumes no alignment. We expect alignment everywhere because the input densities do not exceed $10^{12}$\,cm$^{-3}$.  As noted in \citet{Padovani2012}, while a maximum grain alignment efficiency of 15\% leads to values of the maximum polarization fraction that are consistent with observations, it is still fundamentally an arbitrary parameter.  For both our edge-on and face-on simulations, we have explored the dependence of maximum polarization fraction on the maximum grain alignment efficiency (see Figure \ref{fig:maxp} in the Appendix).

\texttt{LIME} returns three image slices corresponding to the Stokes $I$, $Q,$ and $U$ parameters (the total dust continuum emission and linearly polarized components, respectively). We reconstruct these parameters into a Stokes $I$ (total dust emission) contour map and an array of polarization orientations with polarization angle, $\chi$, defined as

\begin{equation}
\chi= \frac{1}{2} \arctan\left(\frac{U}{Q}\right)
\label{eqn:theta}
\end{equation}

\noindent
We define the magnetic field orientation to be perpendicular to the polarization orientation \citep{Lazarian2011}.

\subsection{Synthetic CO Observations}
\label{sec:synCO}

We complement our outflow identification by producing synthetic maps of $^{12}$CO(2--1), a line commonly used to identify molecular outflows. We use {\sc radmc-3d}\footnote{\url{http://www.ita.uni-heidelberg.de/~dullemond/software/radmc-3d/}} to perform the line radiative transfer. We adopt the non-LTE large velocity gradient (LVG) approximation and use the molecular excitation and collisional data from the Leiden atomic and molecular database \citep{Schoier2005}.

For the radiative transfer step, we flatten the AMR data to a uniform 256$^3$ grid for a region of 0.065 pc ($\Delta x=$52 AU). To convert the simulation mass densities to CO densities, we adopt an abundance of $10^{-4}$ CO per H$_2$ \citep[e.g.,][]{OffnerArce2014}. The CO abundance for gas with temperatures $>800$ K is set to zero to account for CO dissociation in the ionized jet. The CO abundance is also assumed to be zero for densities $n_{\rm H_2}>2 \times 10^4 ~{\rm  cm}^{-3}$ to account for CO freeze-out onto dust grains in cold dense gas. The spectral cube velocities span $\pm 10$ km/s with a channel width of $dv=$0.08 km~s$^{-1}$.

\section{Results}
\label{sec:results}

In order to compare the alignment of outflows and magnetic fields, we consider only the portion of the simulation data that corresponds to the youngest class of YSOs, namely, Class 0 objects \citep{1984ApJ...287..610L}. These are defined as protostars with spectral energy distributions that peak at sub-millimeter wavelengths ($\lambda > 20~\mu$m) and whose total mass is expected to be dominated by the envelope (i.e., $M_{\textrm{env}} >> M_{\textrm{protostar}}$) \citep{1984ApJ...287..610L, Wilking1989}). However, the protostellar mass cannot be measured directly and the SED classification depends strongly on the viewing angle \citep{robitaille07,Offner12}. Class 0 objects are associated with the main accretion phase, exhibit strong and centrally condensed dust continuum emission, and have clear bipolar outflows \citep{Andre2000, 2014prpl.conf..195D}. The last of these was a selection criterion of the Class 0 sources observed by \citet{Hull2014}. Since the Class 0 lifetime is estimated to last $\sim$0.1 Myr \citep{evans09}, these sources are on average very young.

\begin{figure}
\centering
\includegraphics[width=1.0\linewidth, clip, trim=0.5cm 0.0cm 2.0cm 0.0cm]{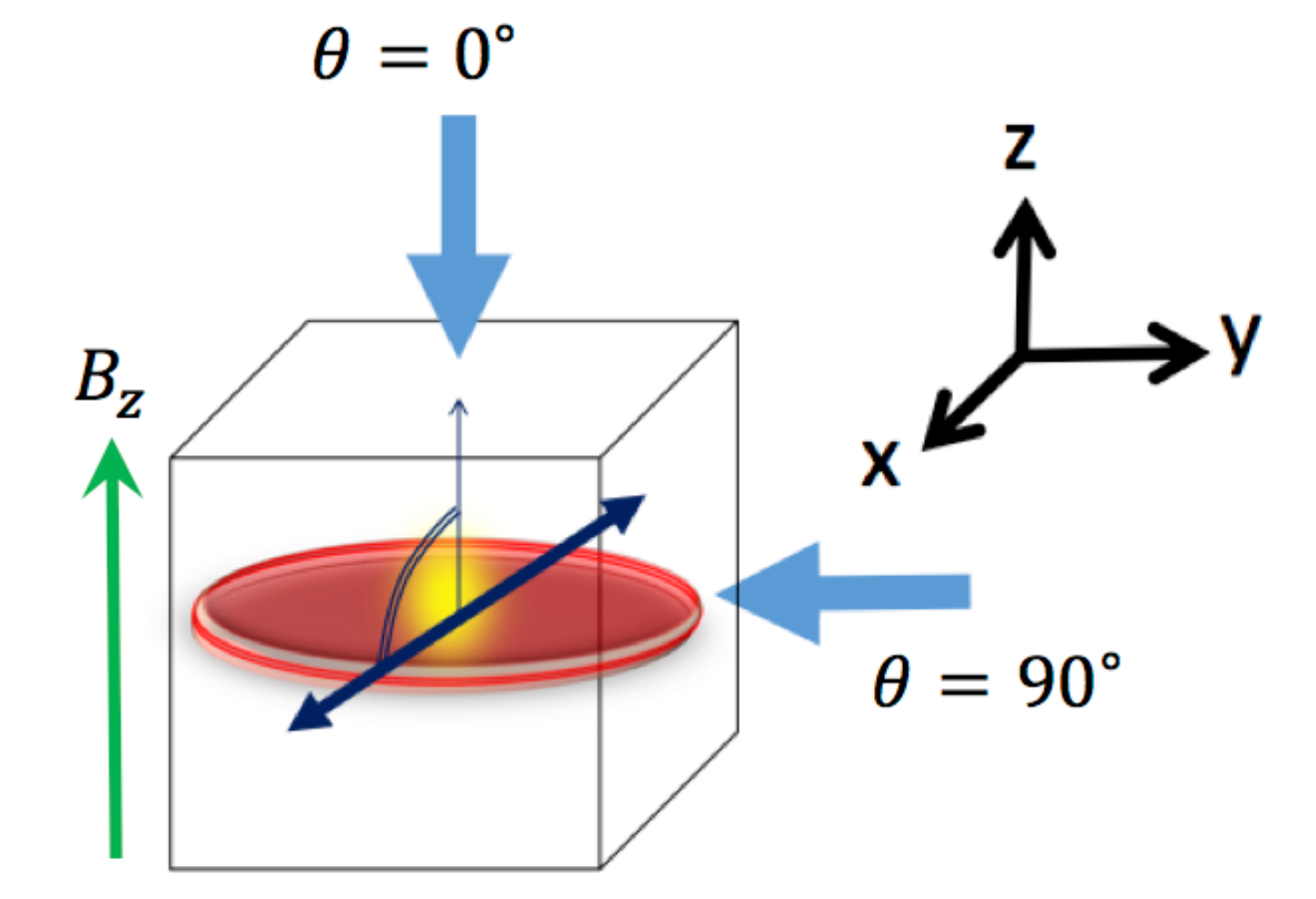}
\caption{\footnotesize
A schematic of a collapsing protostellar core. Blue arrows: face-on (0$\degree$) and edge-on views (90$\degree$); green arrow: initial magnetic field direction. Cube interior: a representation of the collapsing, flattened core,  
with double-headed arrow showing the initial outflow orientation.}
\label{fig:schematic}
\end{figure}

\begin{figure*}[hbt!]
\centering
\includegraphics[width=0.49\linewidth, clip, trim=0.3cm 0.0cm 1.0cm 1.2cm]{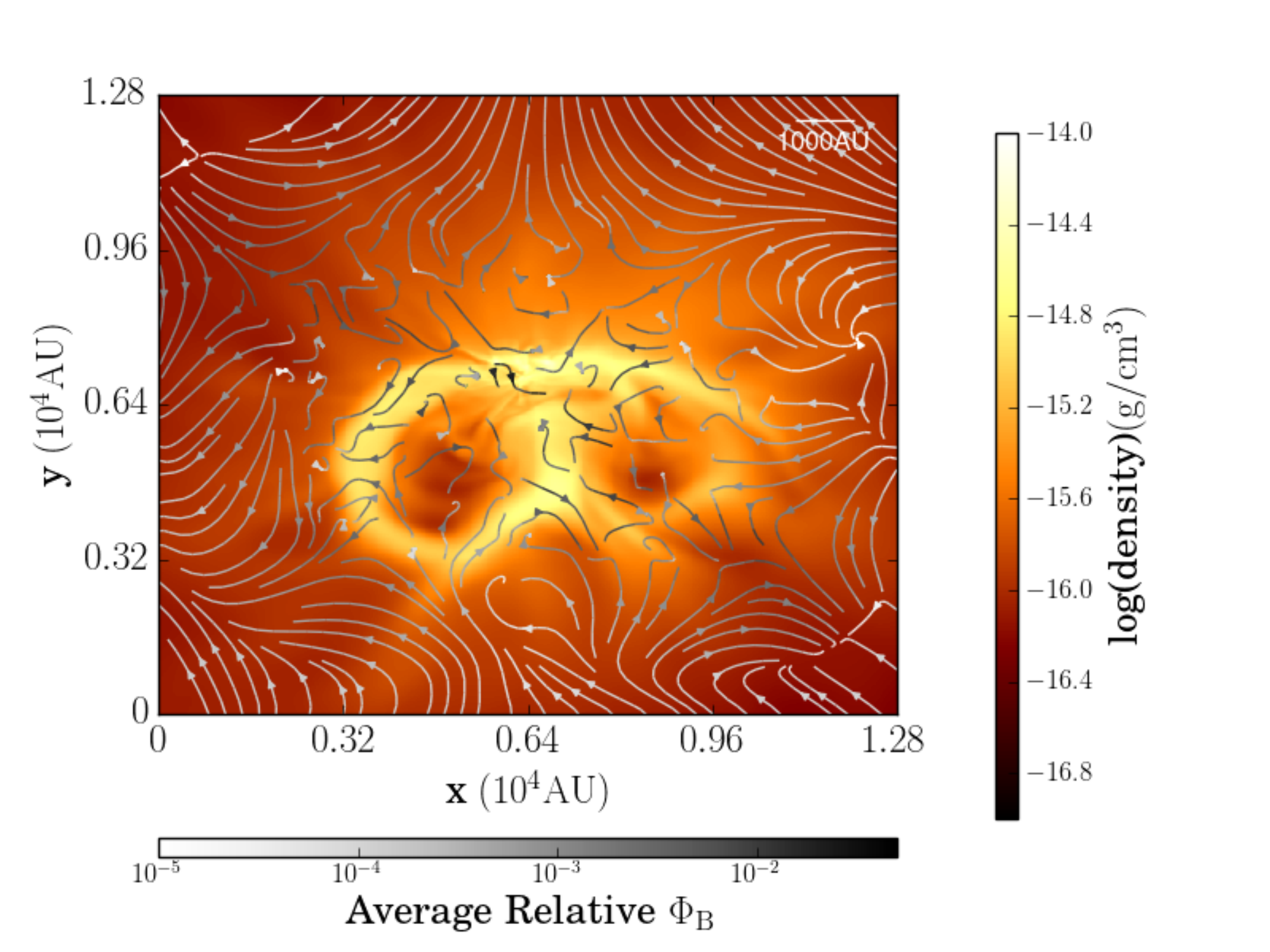}
\includegraphics[width=0.49\linewidth, clip, trim=0.3cm 0.0cm 1.0cm 1.2cm]{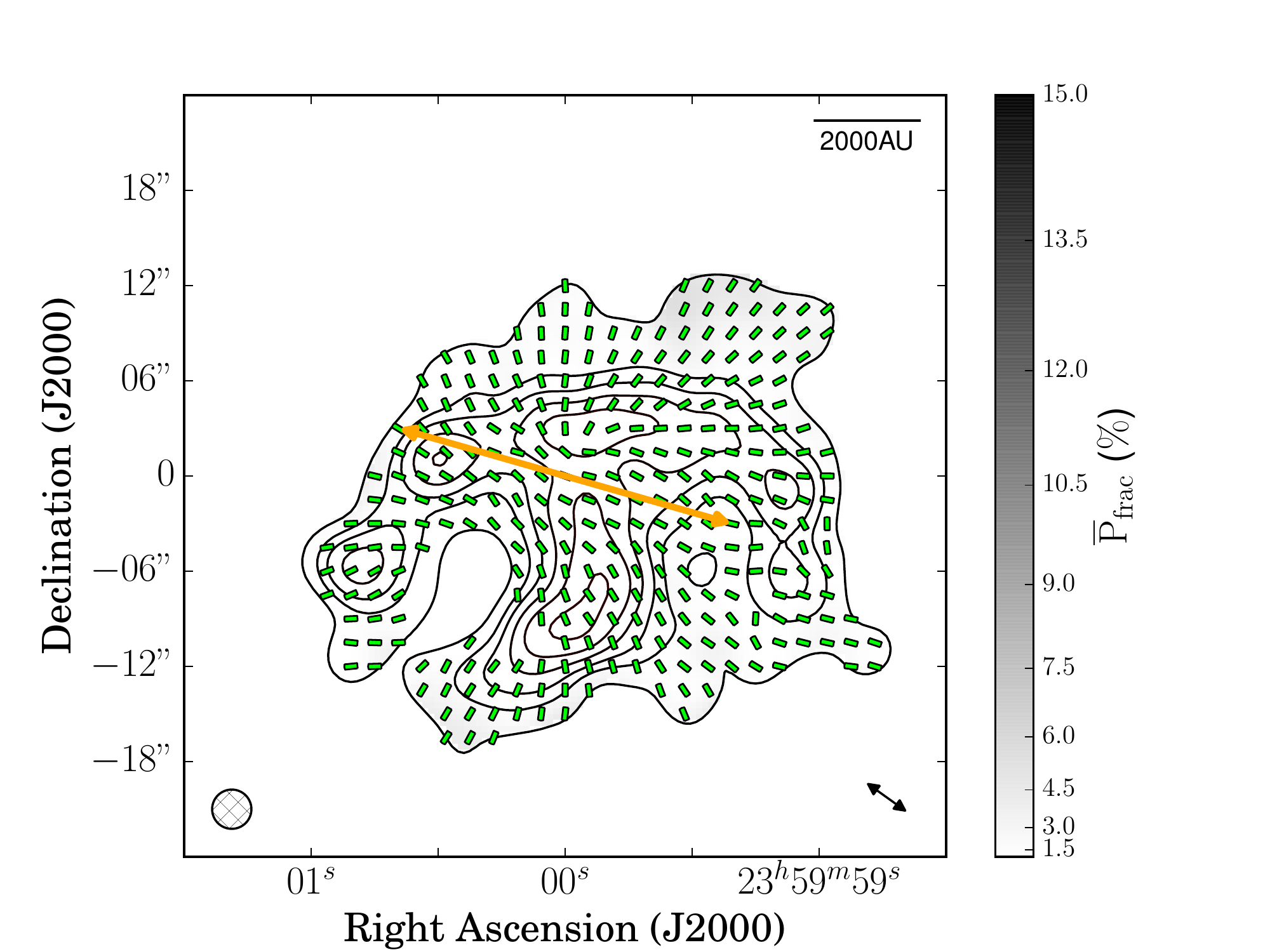}
\includegraphics[width=0.49\linewidth, clip, trim=0.3cm 0.0cm 1.0cm 1.2cm]{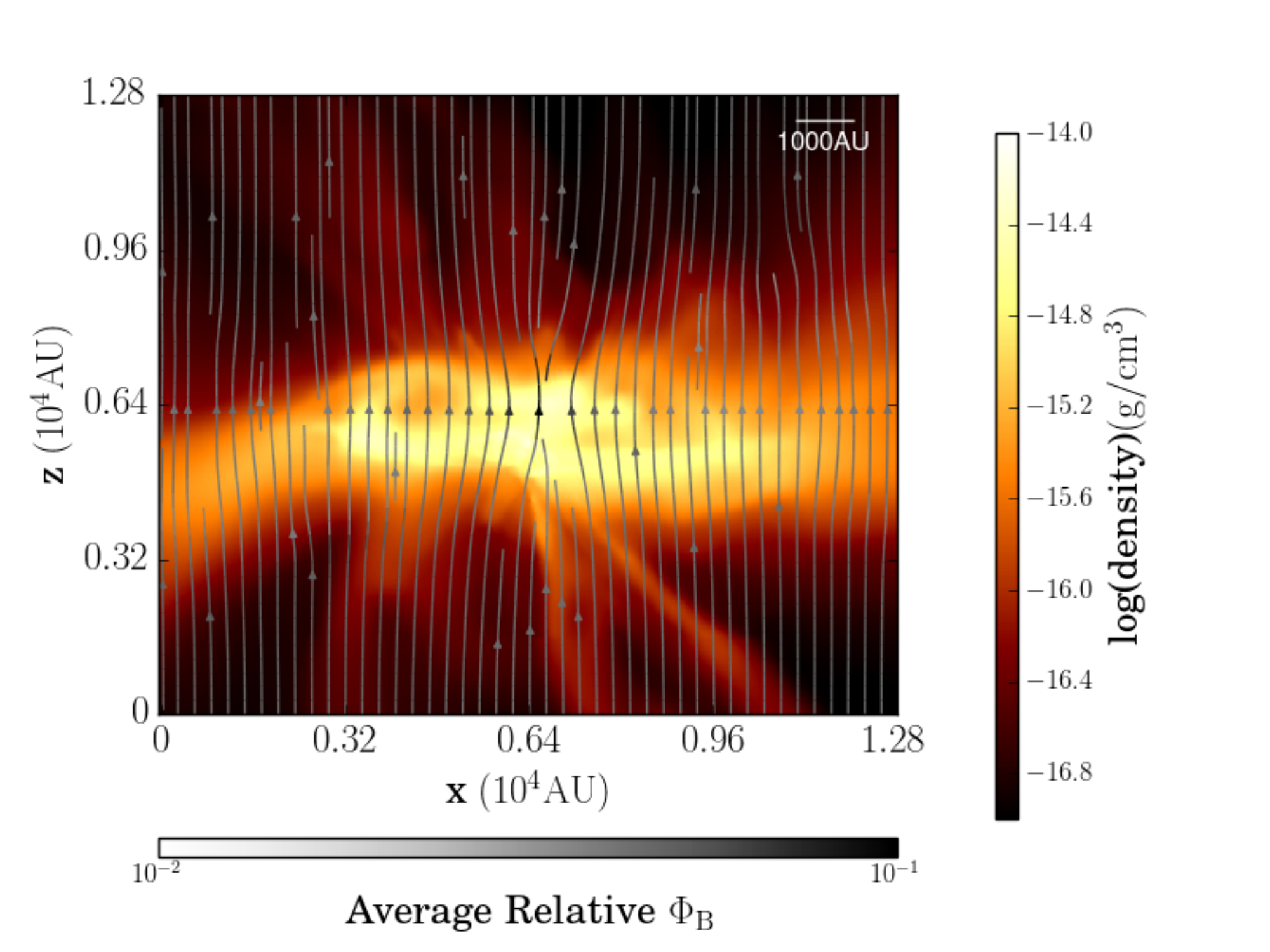}
\includegraphics[width=0.49\linewidth, clip, trim=0.3cm 0.0cm 1.0cm 1.2cm]{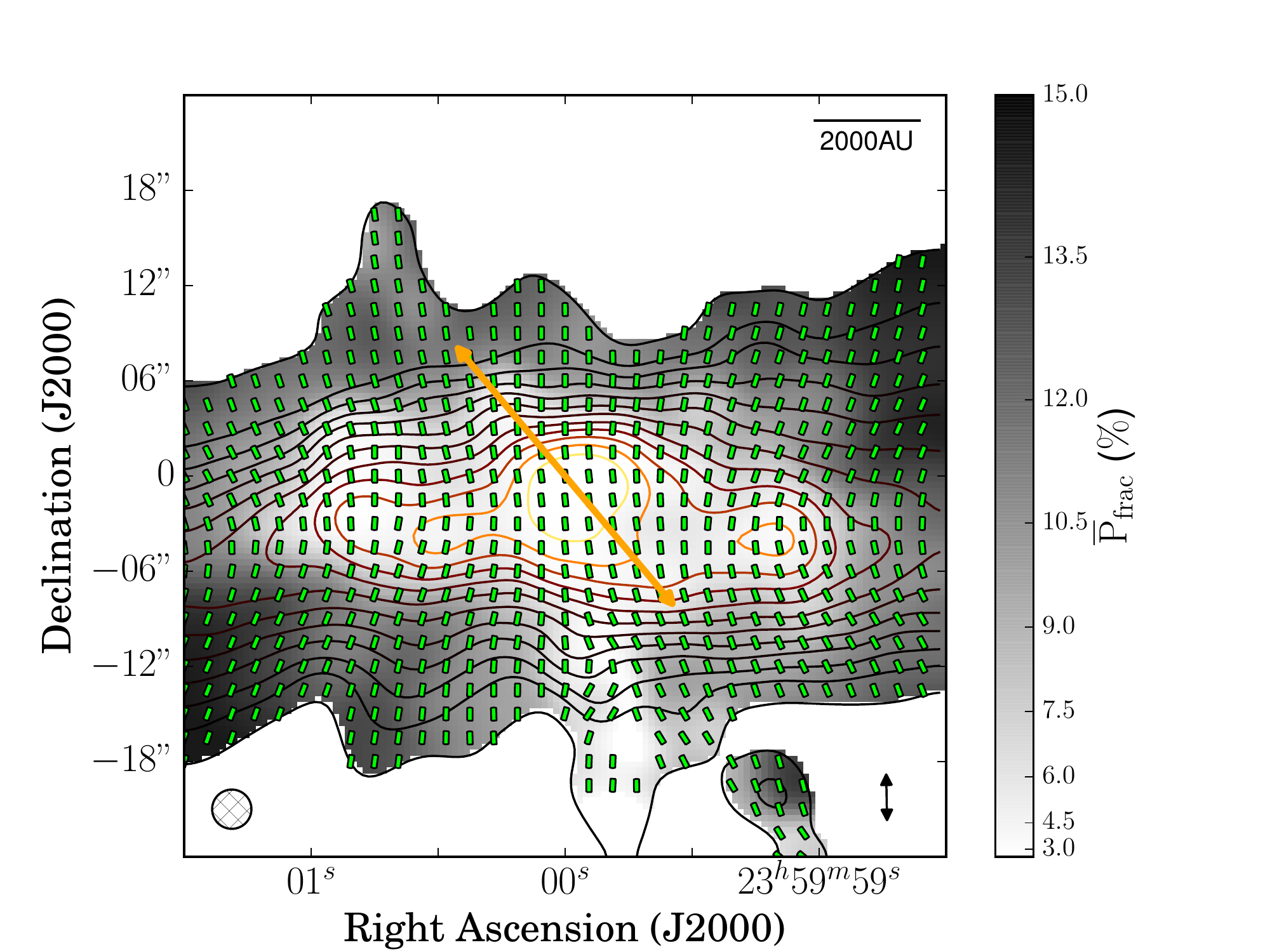}
\includegraphics[width=0.49\linewidth, clip, trim=0.3cm 0.0cm 1.0cm 1.2cm]{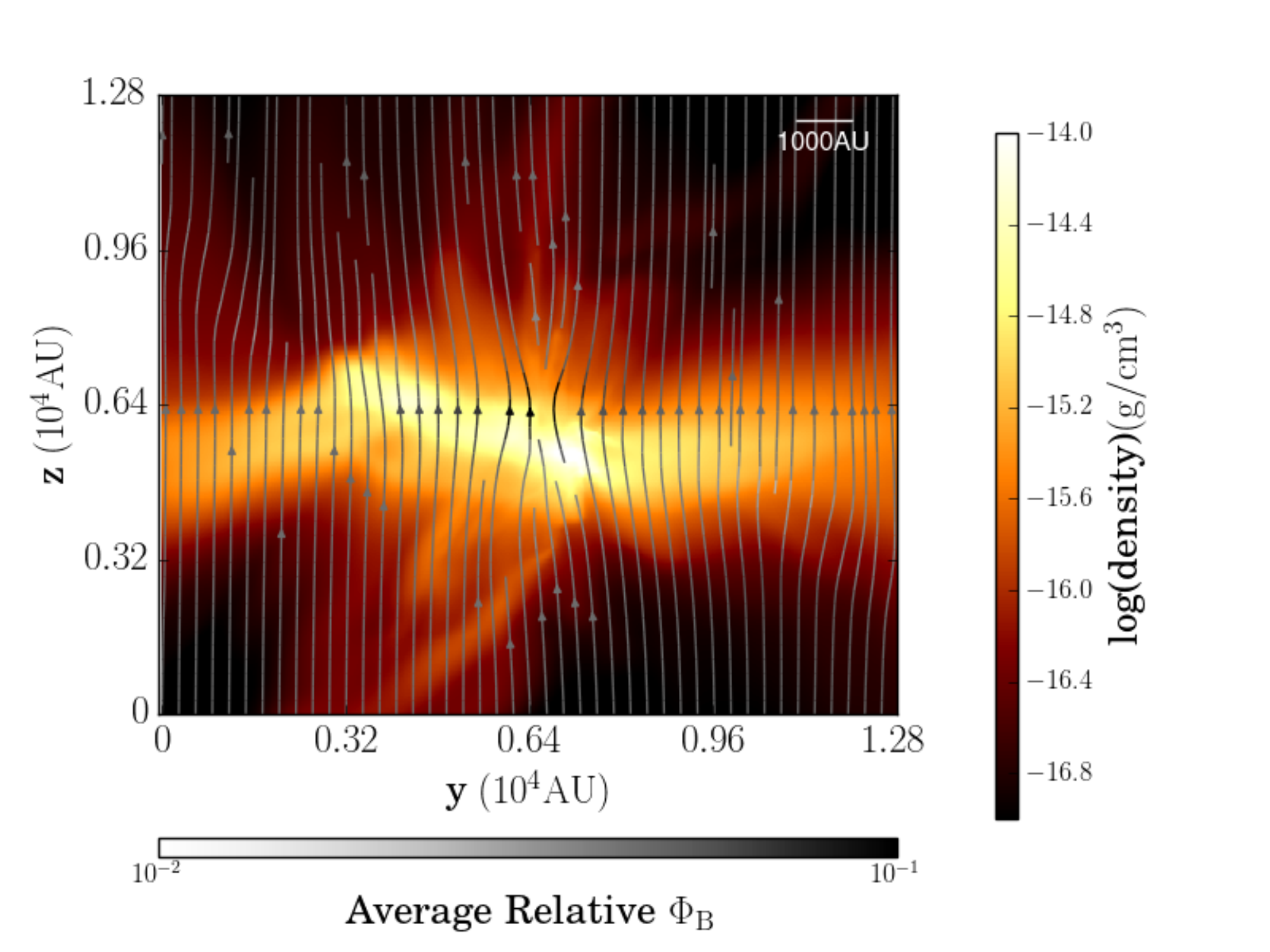}
\includegraphics[width=0.49\linewidth, clip, trim=0.3cm 0.0cm 1.0cm 1.2cm]{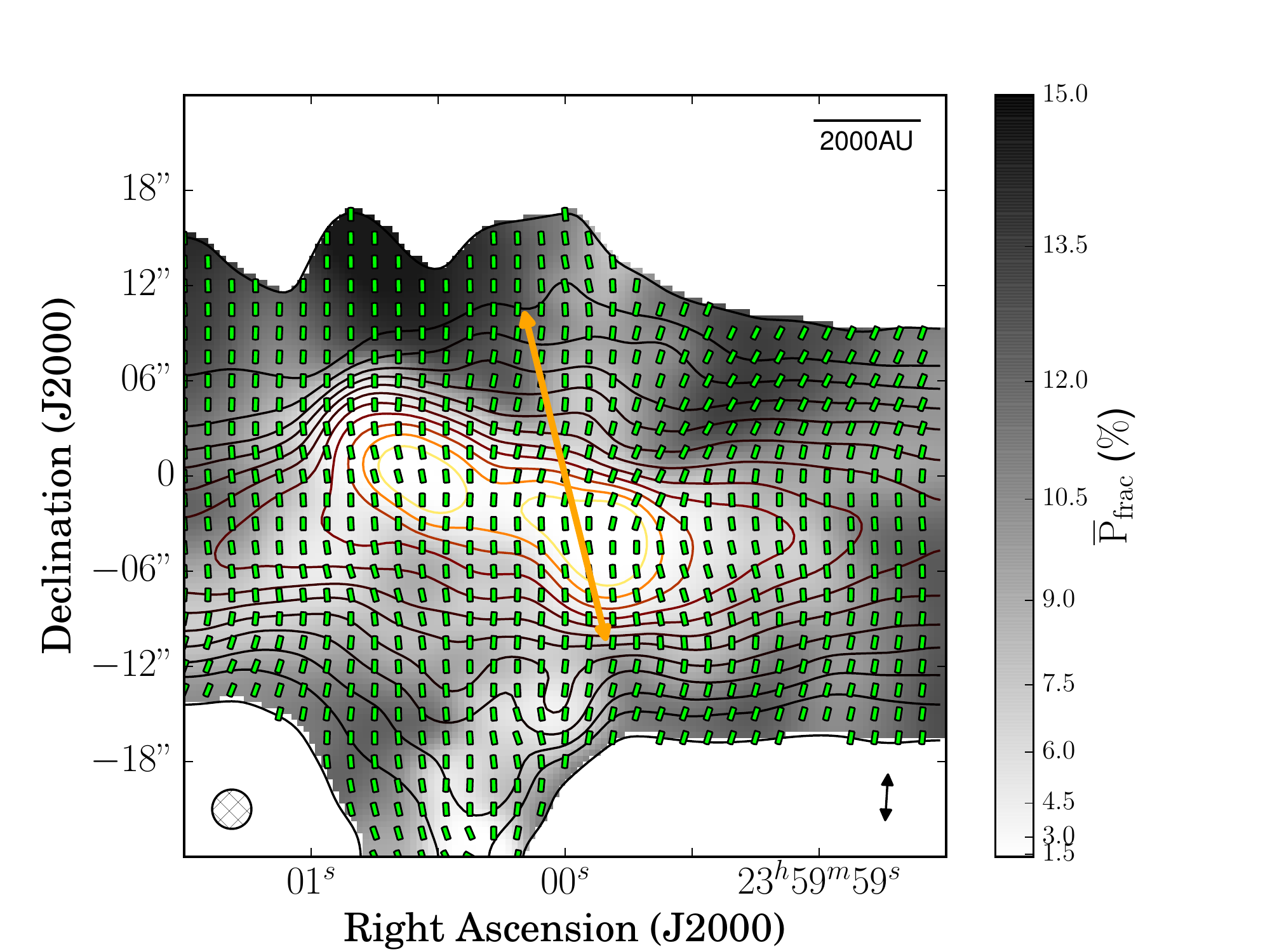}
\caption{\footnotesize
Visualizations of column density and magnetic field \textit{(left)} and synthetic observations of dust polarization \textit{(right)} of the S1.5 simulation at $t=0.353$\,Myr.
\textit{Top:} face-on view from the $z$-direction, where $\theta=0\degree$. 
\textit{Middle:} edge-on view from the $y$-direction, where $\theta=90\degree$ and $\phi = 0\degree$.
\textit{Bottom:} edge-on view from the $x$-direction, where $\theta=90\degree$ and $\phi = 90\degree$.
Contours: Stokes $I$ thermal dust emission at 3, 5, 7, 10, 14, 20, 28, 40, 56, 79, 111, 155, and 217 $\times \sigma_{I}$ where $\sigma_{I}$ is the rms noise in Stokes $I$. Green line segments: magnetic field orientation (perpendicular to polarization orientation). Grayscale background: polarization fraction $\bar{P}_{\textrm{frac}}$ (\%) shown with a power law ($\gamma=2$). The orange double-headed arrow shows the outflow orientation.  The small, black, double-headed arrow in the bottom-right of each synthetic observation (right-hand panels) indicates the average (unweighted) magnetic field orientation.
}
\vspace{1in}
\label{fig:Dustpol}
\end{figure*}

We render the visualizations and synthetic observations for each time step in the simulations from a number of different viewing angles. The face-on and edge-on views of the system are $\theta=0\degree$ and $\theta=90\degree$, respectively (see Figure~\ref{fig:schematic} for a schematic of the viewing orientations). Figure~\ref{fig:Dustpol} shows examples of face-on (top row) and edge-on views (middle and bottom rows) of S1.5 at $t=0.353$\,Myr. The right panels show synthetic thermal dust polarization and emission maps (see \S\ref{sec:method3}). The corresponding column density and projected magnetic field maps are on the left.  The magnetic field orientations from the synthetic dust polarization maps are consistent with the magnetic field in the raw simulation visualizations. Features in column density also appear in the Stokes $I$ maps from \texttt{DustPol}.

In the top row of Figure~\ref{fig:Dustpol}, which shows the face-on view of the magnetic field, there is no clear overall magnetic field orientation. In contrast, edge-on views from the $y$- and $x$-directions (middle and bottom rows of Figure~\ref{fig:Dustpol}, respectively) show that the magnetic field morphology is vertical when viewed from both of these orthogonal edge-on views.  We have produced additional edge-on synthetic observations using intermediate values of $\theta$ (not pictured here). Due to the initial azimuthal symmetry of the simulation and the fact that the initial magnetic field is in the $z$-direction, we find that all edge-on views look very similar. 

There are also hints of the characteristic ``hourglass'' morphology in the edge-on views. This is similar to observations of hourglass-shaped fields in sources such as NGC 1333-IRAS 4A \citep{Girart2006} and L1157 \citep{Stephens2013, Hull2014}, where gravitational collapse of a strongly magnetized protostellar core is expected to pinch magnetic field lines.

\subsection{Polarization Fraction}
\label{polfrac} 
 
\begin{figure*}[hbt!]
\centering
\includegraphics[width=0.43\linewidth, clip, trim=0.3cm 0.0cm 1.0cm 1.2cm]{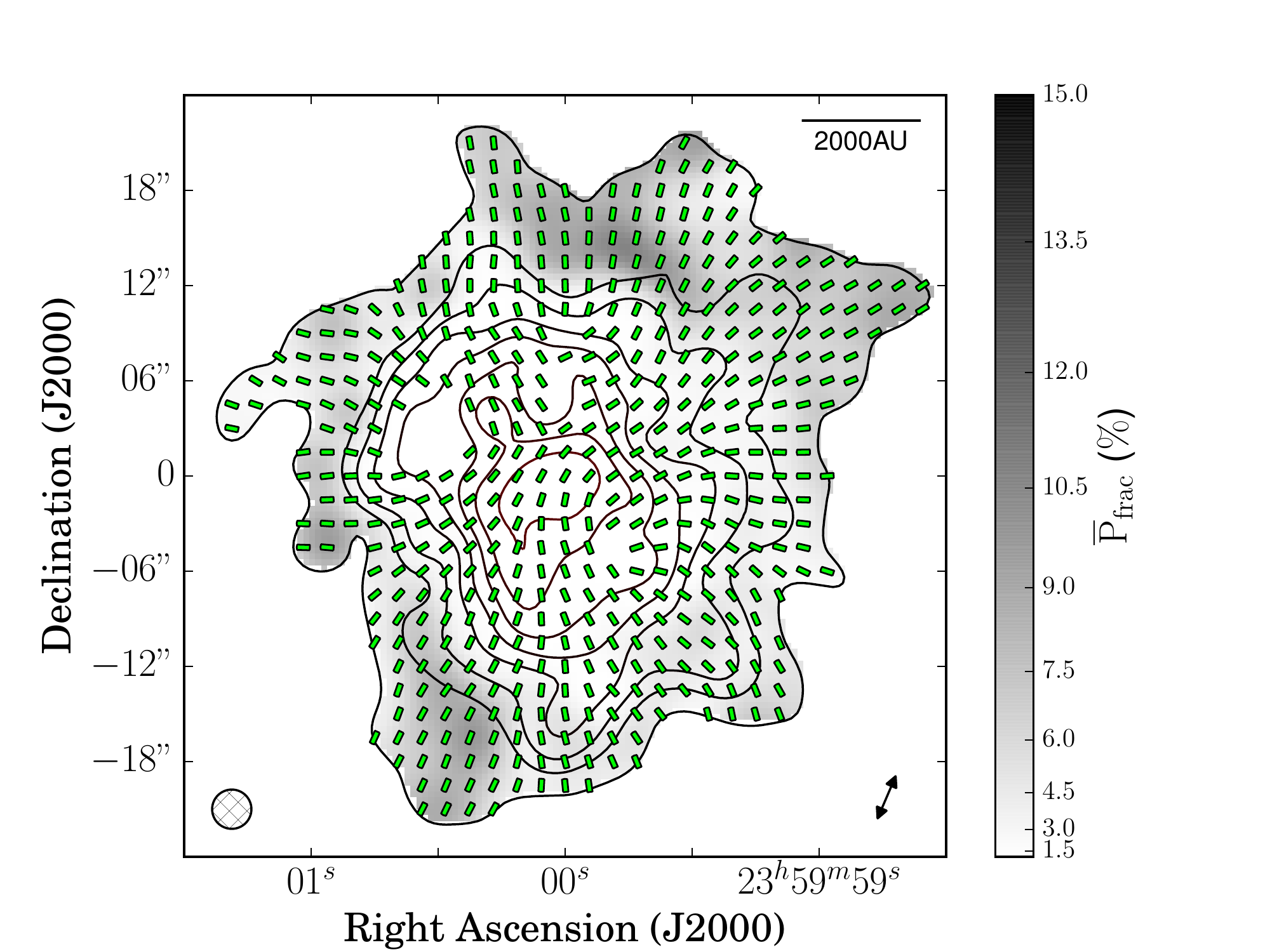}
\includegraphics[width=0.4\linewidth, trim=0.3cm 5.6cm 1.0cm 5.6cm]{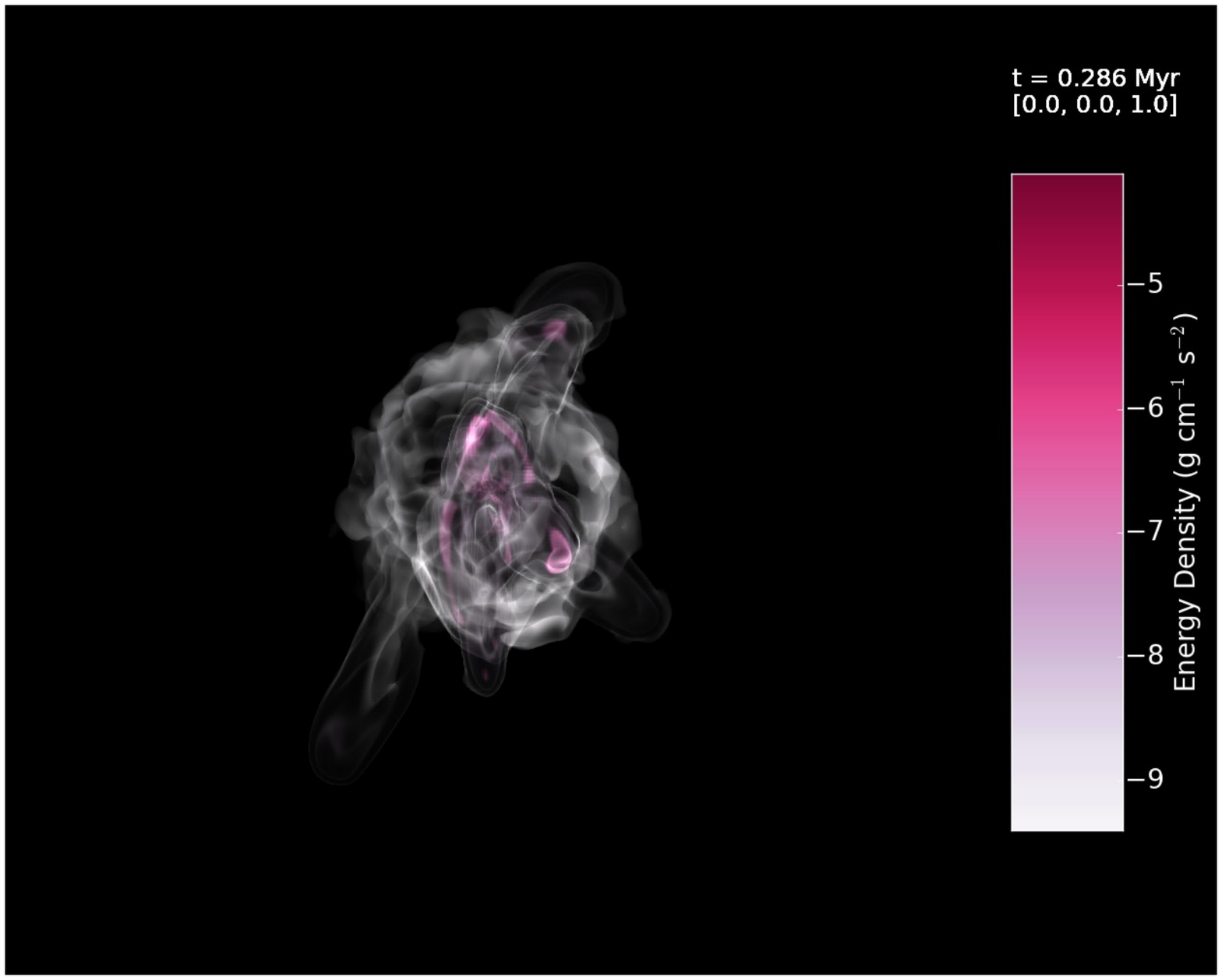}
\includegraphics[width=0.43\linewidth, clip, trim=0.3cm 0.0cm 1.0cm 1.2cm]{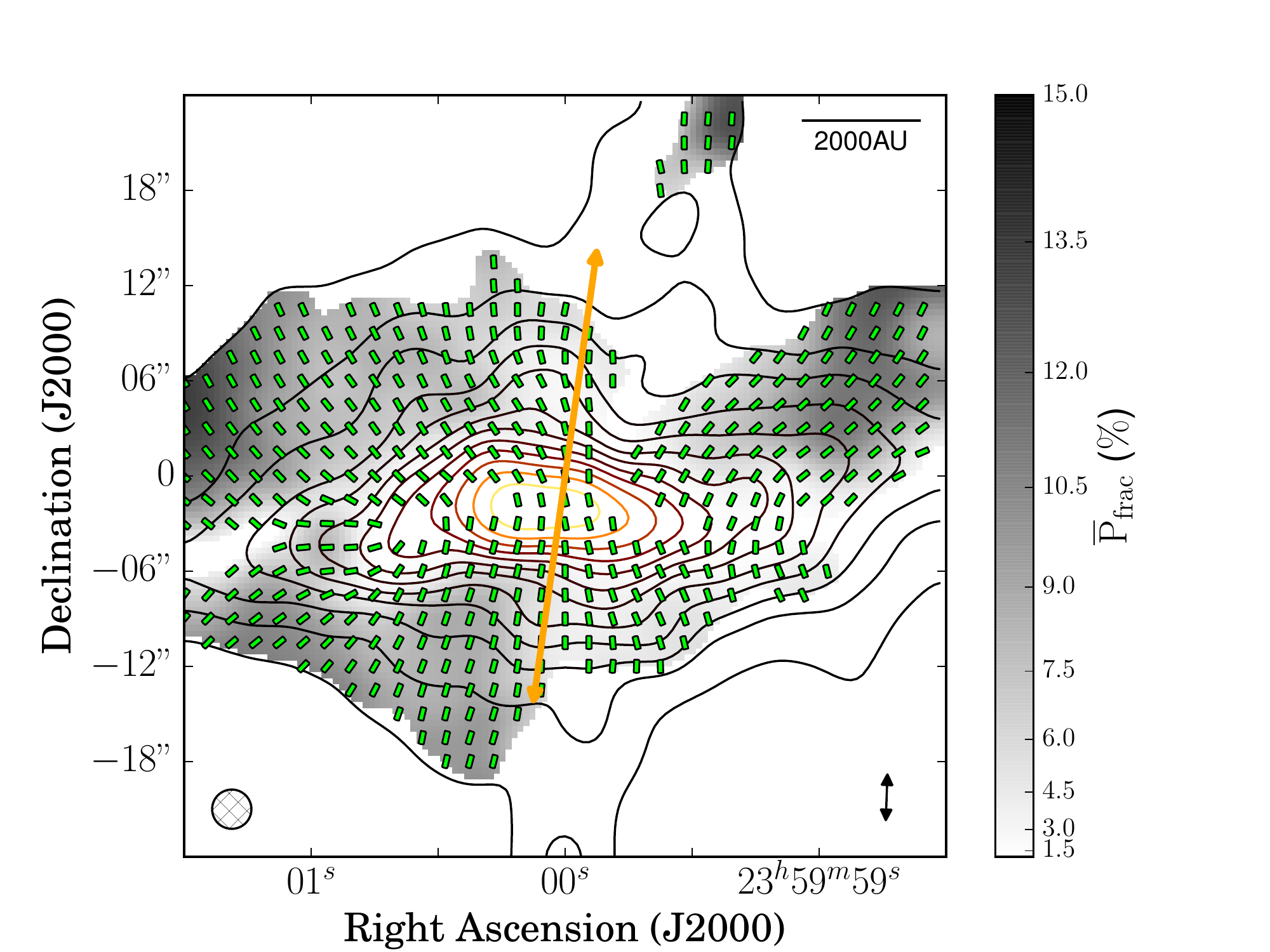}
\includegraphics[width=0.4\linewidth, clip, trim=0.3cm 5.6cm 1.0cm 5.6cm]{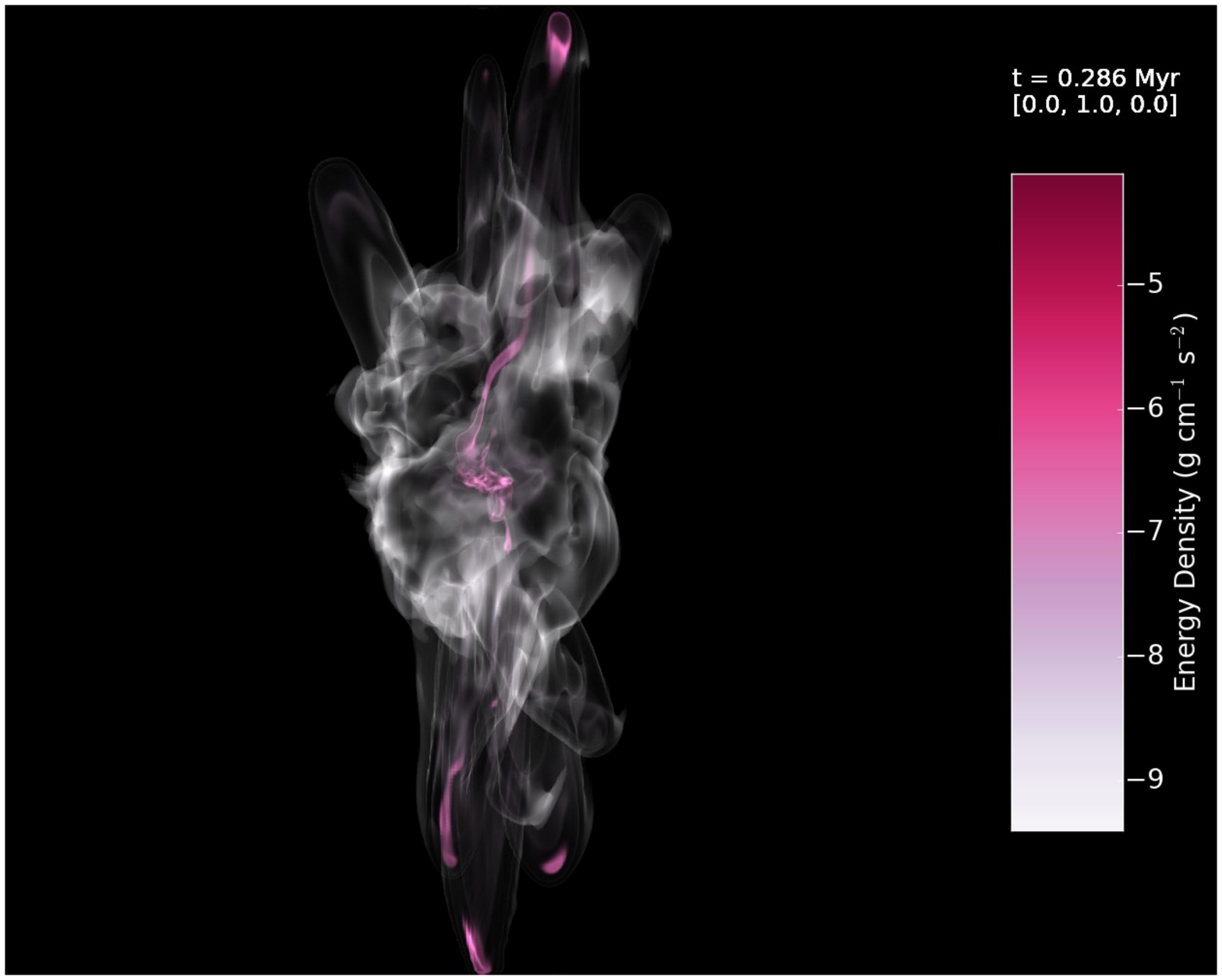}
\includegraphics[width=0.43\linewidth, clip, trim=0.3cm 0.0cm 1.0cm 1.2cm]{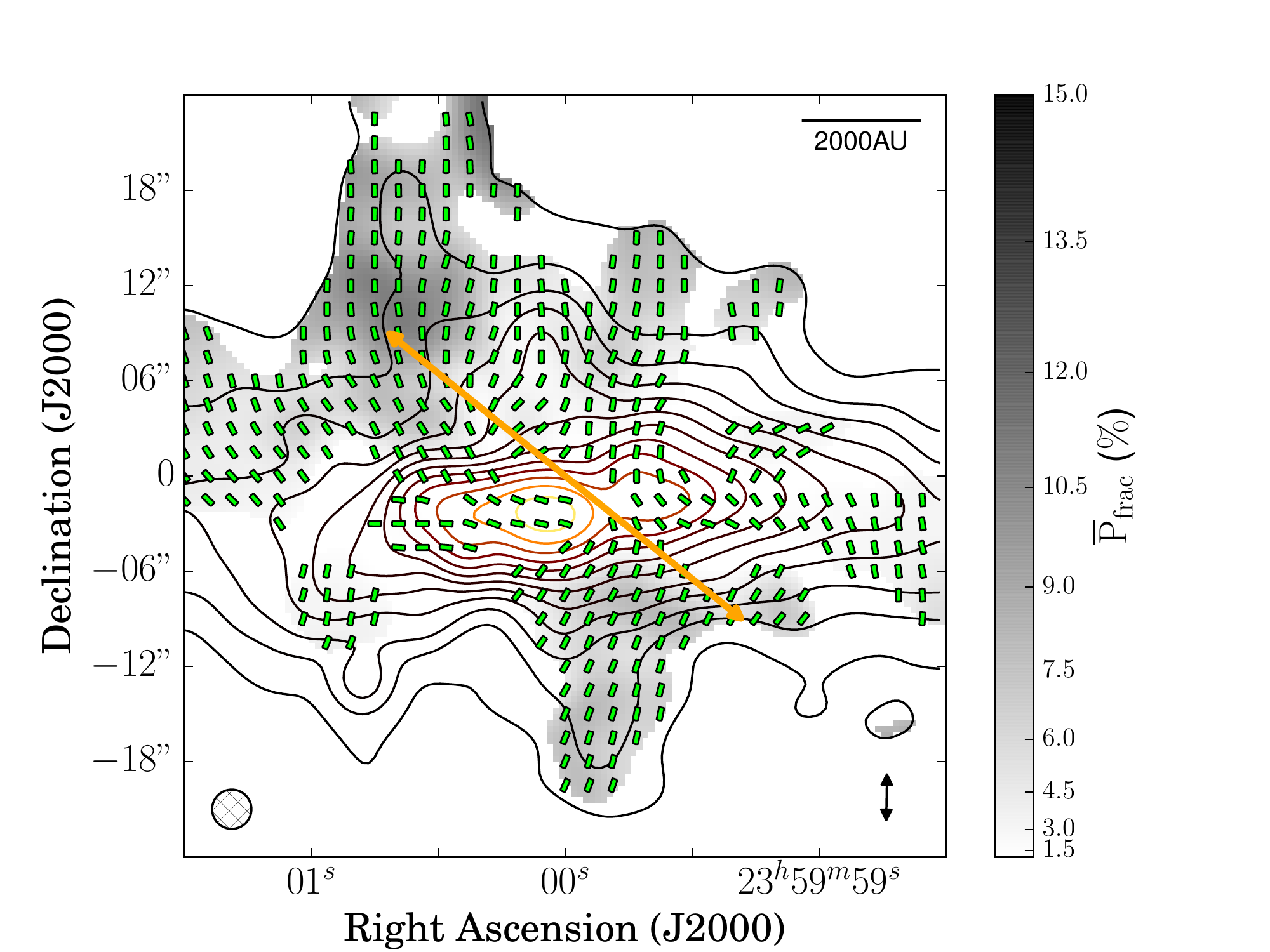}
\includegraphics[width=0.4\linewidth, clip, trim=0.3cm 5.6cm 1.0cm 5.6cm]{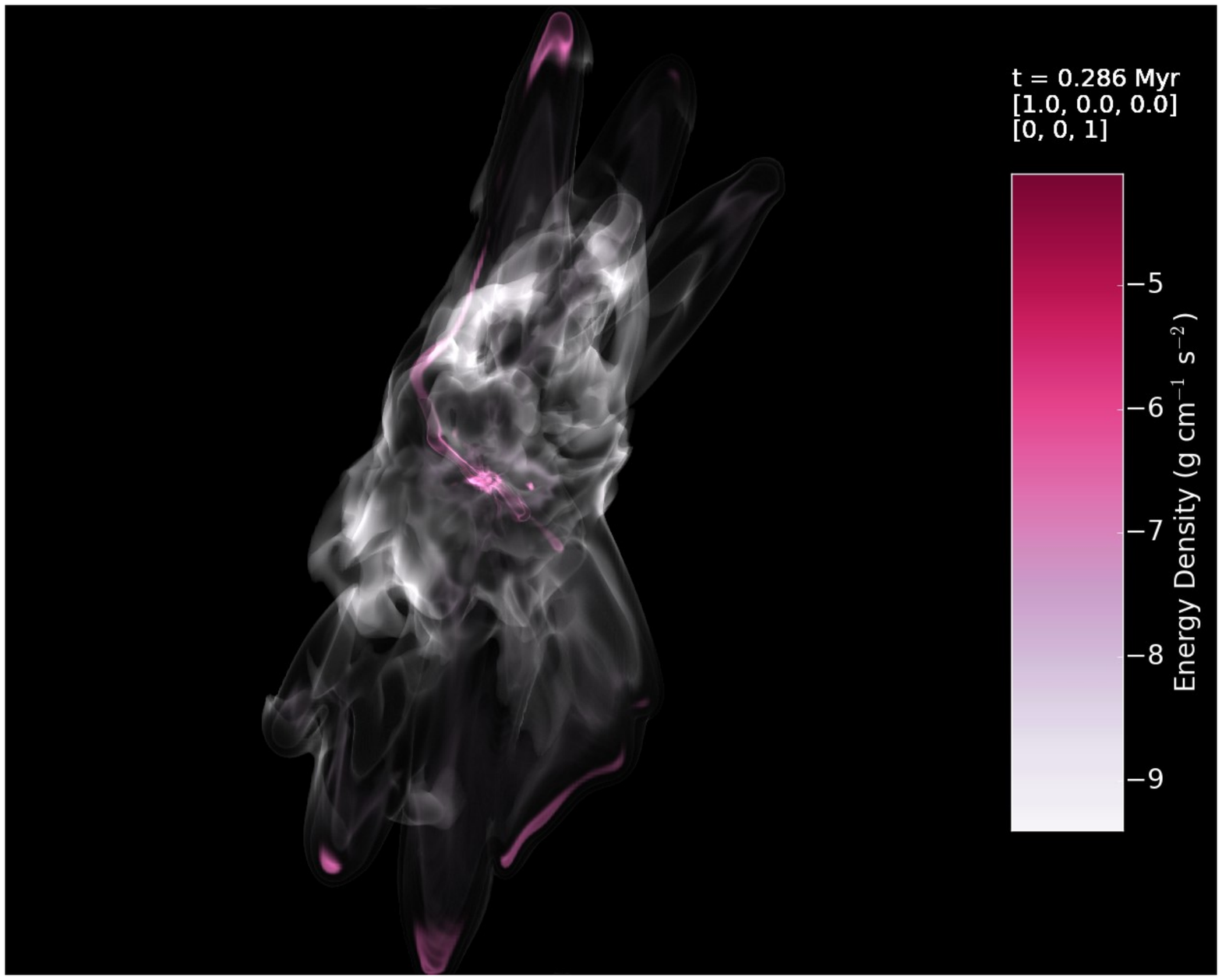}
\caption{\footnotesize
From top to bottom: the $x$-$y$ plane (face-on), $x$-$z$ plane, and $y$-$z$ plane (both edge-on) viewed from $z, y,$ and $x$ directions respectively. \textit{Left:} synthetic dust polarization maps of S2.5 at $t=0.286$ Myr, plotted in the same manner as the right-hand panels in Figure \ref{fig:Dustpol}.  The outflow direction was not discernible in the face-on view. \textit{Right:} the corresponding energy density, where high energy densities in pink indicate the location of the outflow.} 
\vspace{0.2 in}
\label{fig:Energy density xyz}
\end{figure*}

\begin{figure*} [hbt!]
\centering
\includegraphics[width=0.43\linewidth, clip, trim=0.3cm 0.0cm 1.0cm 1.2cm]{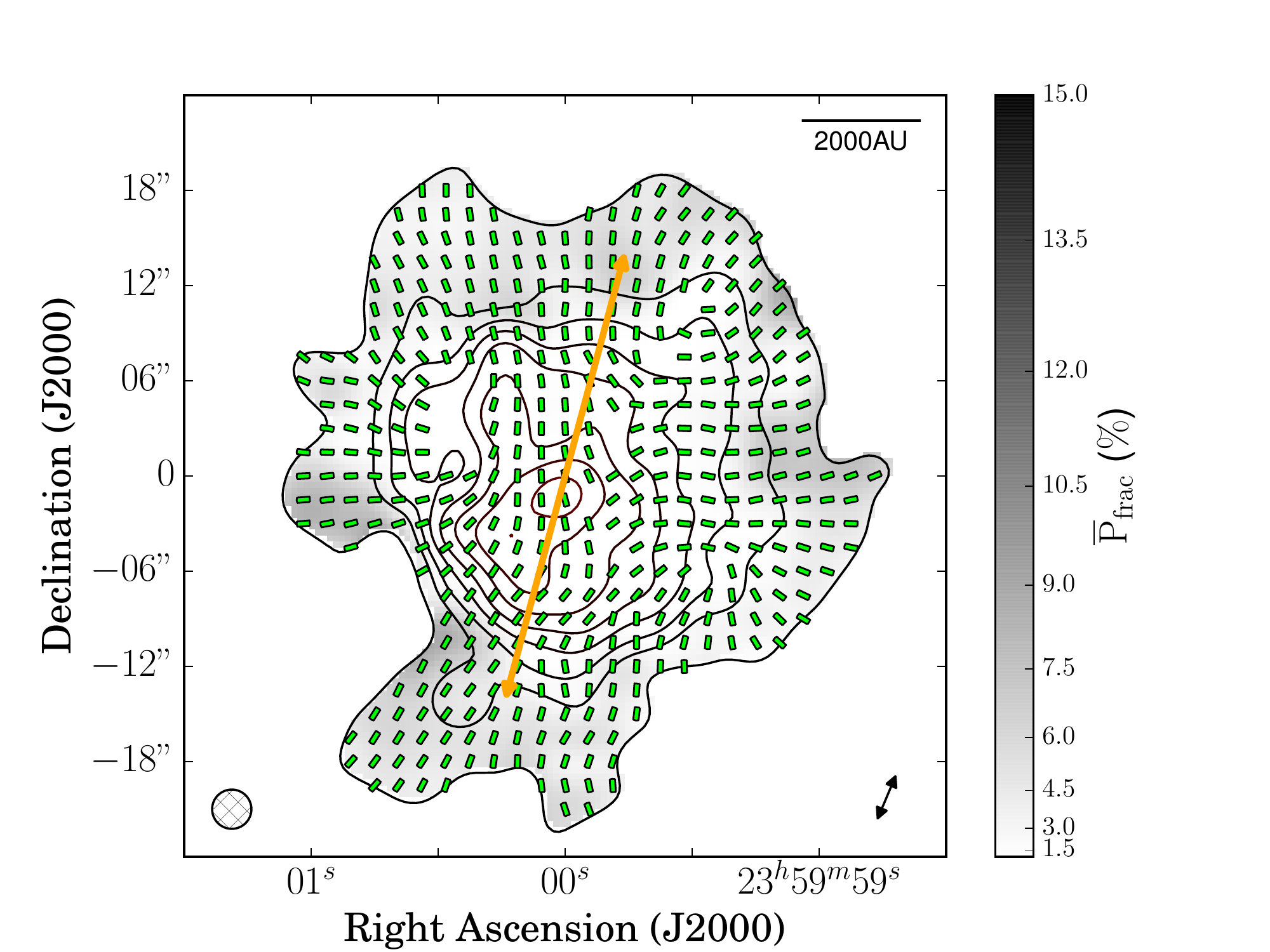}
\includegraphics[width=0.4\linewidth, trim=0.3cm 5.6cm 1.0cm 5.6cm]{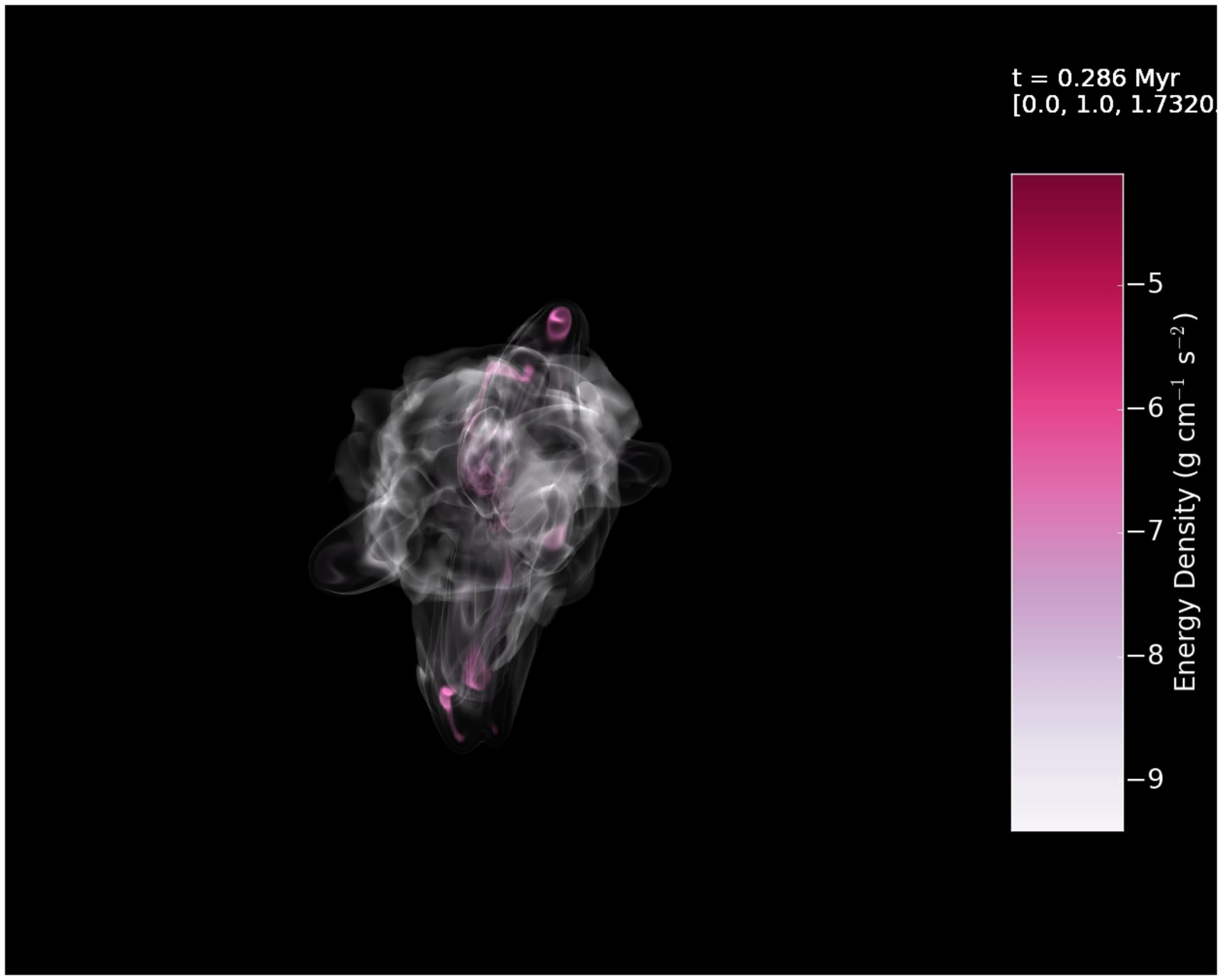}
\includegraphics[width=0.43\linewidth, clip, trim=0.3cm 0.0cm 1.0cm 1.2cm]{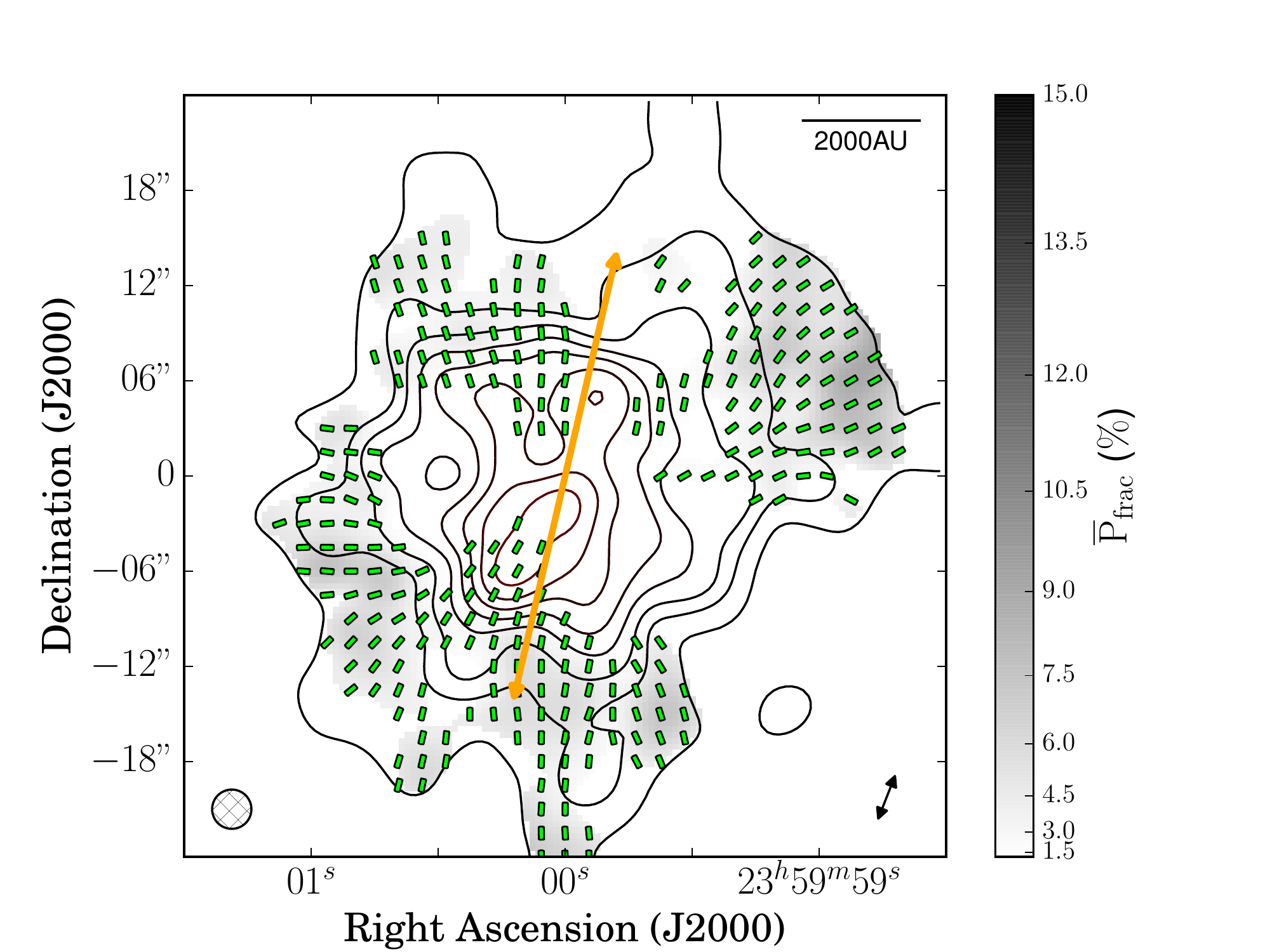}
\includegraphics[width=0.4\linewidth, clip, trim=0.3cm 5.6cm 1.0cm 5.6cm]{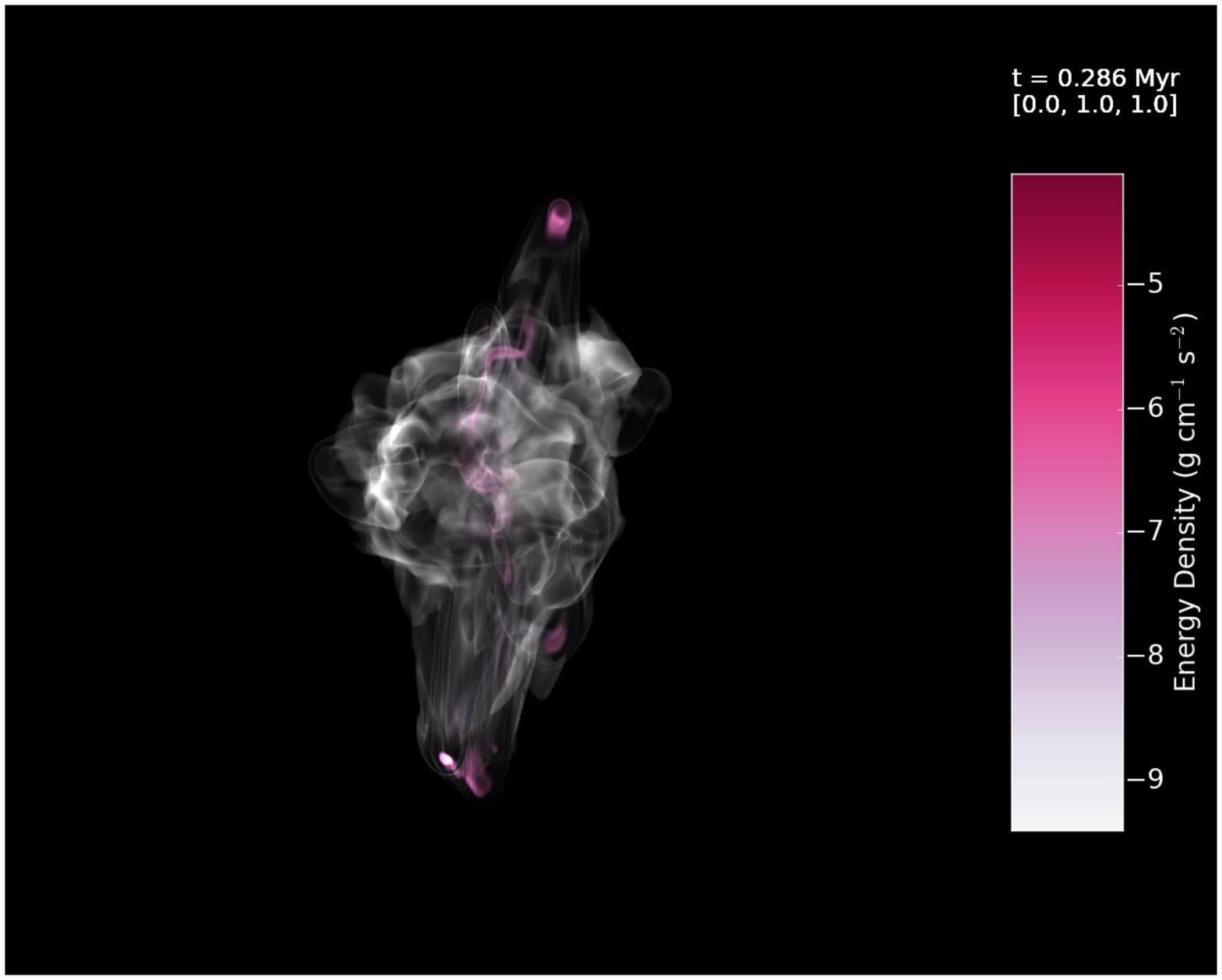}
\includegraphics[width=0.43\linewidth, clip, trim=0.3cm 0.0cm 1.0cm 1.2cm]{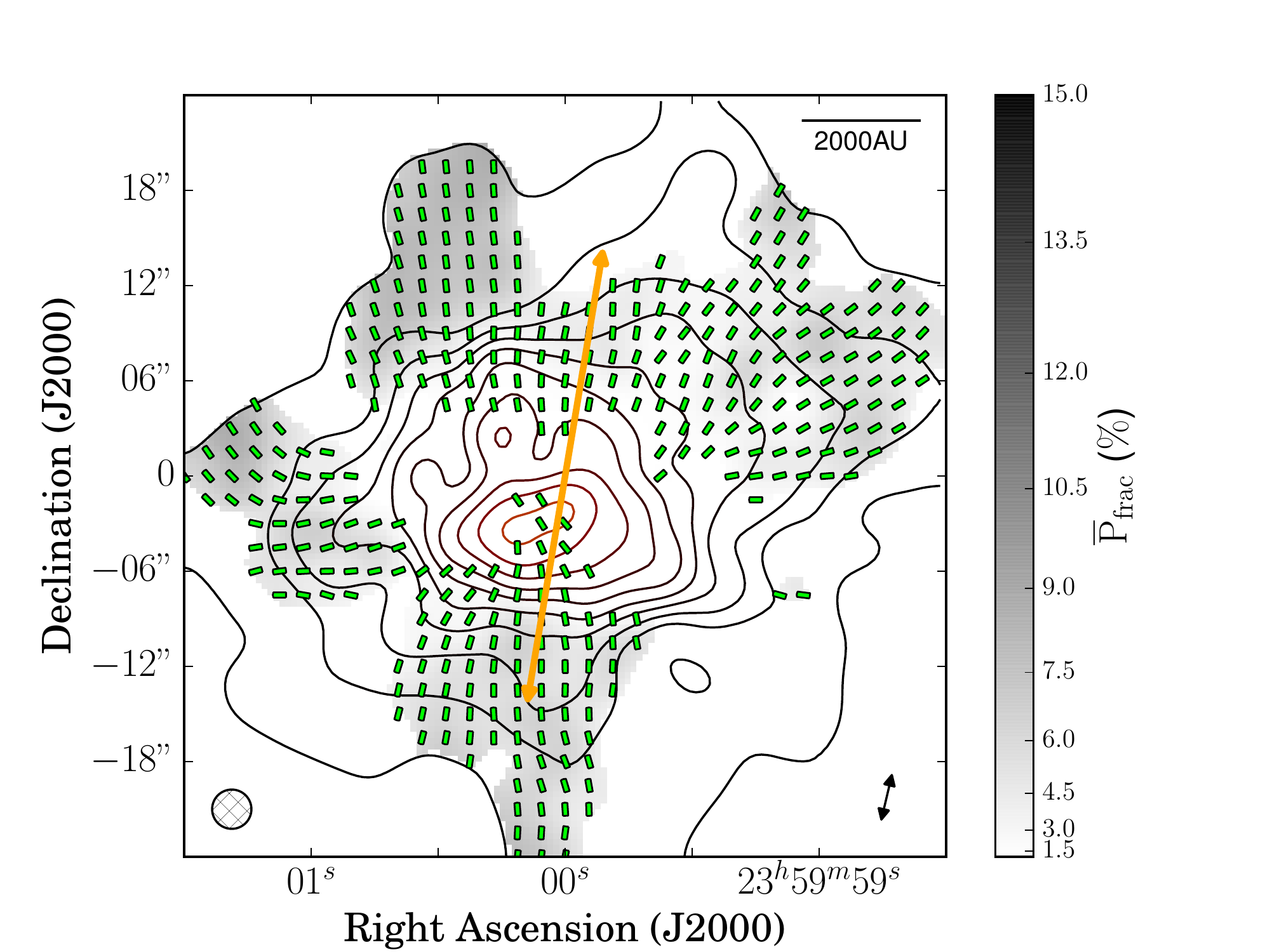}
\includegraphics[width=0.4\linewidth, clip, trim=0.3cm 5.6cm 1.0cm 5.6cm]{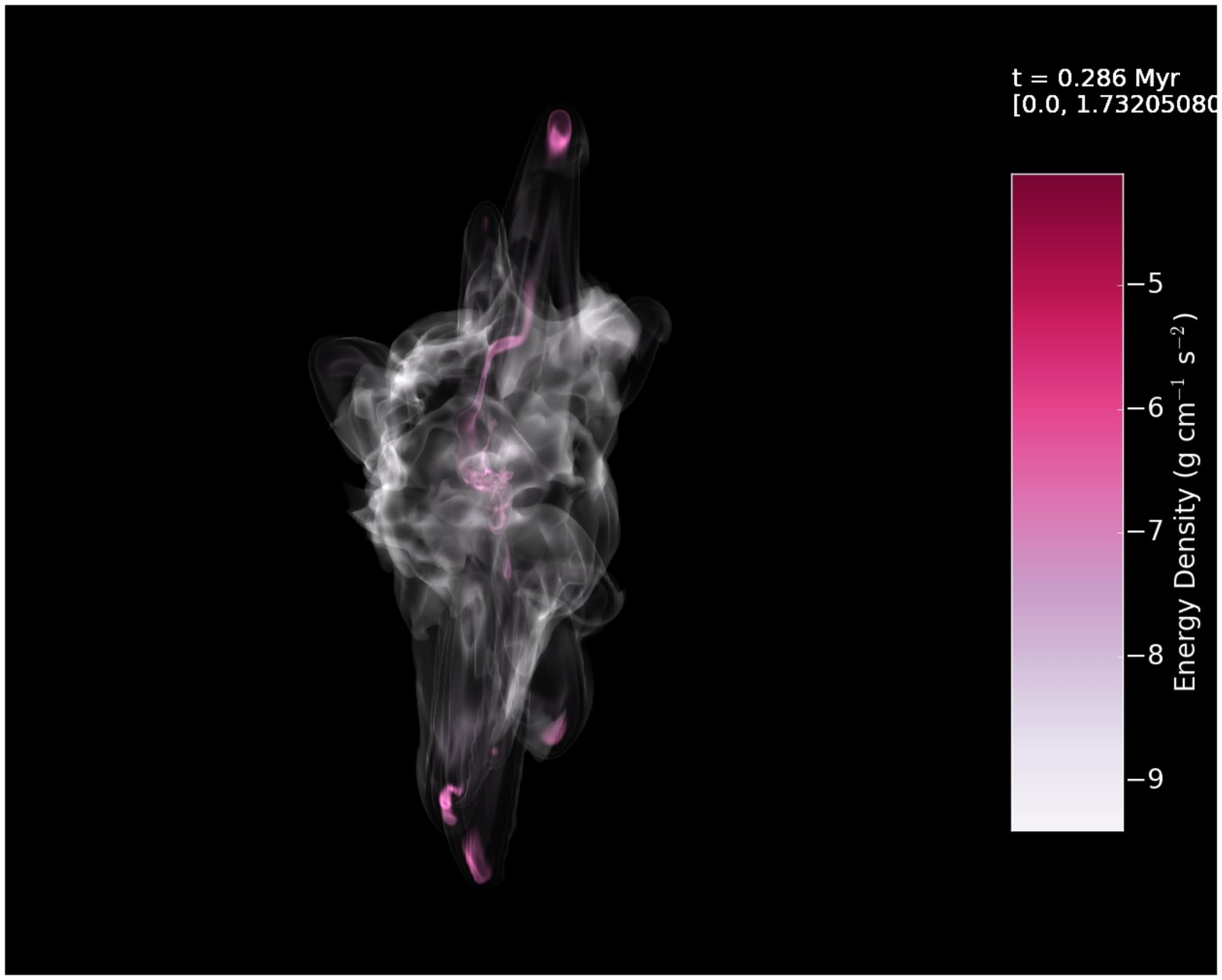}
\caption{\footnotesize 
Same as Figure \ref{fig:Energy density xyz}, but for intermediate viewing angles $30\degree, 45\degree$, and $60\degree$, where where $\theta=0\degree$ corresponds to the face-on view from the $z$-direction, and $\theta = 90\degree$ corresponds to the edge-on view from the $y$-direction. } 
\vspace{0.2in} 
\label{fig:Energy density 304560} 
\end{figure*}

To explore projection effects, we render projected energy density maps and synthetic polarization maps at viewing angles $\theta = 0\degree, 30\degree, 45\degree, 60\degree$ and $90\degree$, where $\theta=0\degree$ corresponds to the face-on view and $\theta = 90\degree$ corresponds to the edge-on view from the $y$-direction. Figure~\ref{fig:Energy density xyz} shows the orthogonal viewing angles along $x,$ $y,$ and $z$. Figure~\ref{fig:Energy density 304560} shows the views from $30\degree$, $45\degree$, and $60\degree$ (top to bottom). We present the synthetic dust polarization maps in the left panel with their corresponding energy density projections on the right.

We calculate the average value of polarization fraction for each dust polarization map by averaging the magnitude of the polarized intensity. The average polarization fraction, $\bar{P}_{\textrm{frac}}$, is sensitive to the weighting scheme used to derive it. Also, the locations of the peak intensities of the total and polarized emission are not the same for every source (this is the case for both our synthetic observations as well as real observations). Therefore, the fractional polarization $\bar{P}_{\textrm{frac}}$ is calculated using the mean (total) intensity, $\bar{I}$, and mean polarized intensity, $\bar{P}$, across the whole source: $\bar{P}_{\textrm{frac}}$ = $\bar{P}/\bar{I}$.  We apply the cutoff conditions $P> 3\,\sigma_{P}$ and $I > 3\,\sigma_{I}$, where $\sigma_{P}$ is the rms noise of the polarized intensity ($\sigma_{P}\approx\sigma_{Q}\approx\sigma_{U}$) and $\sigma_{I}$ is the rms noise of the total intensity.  As is generally the case for real observations, we measure both $\sigma_{P}$ and $\sigma_{I}$ in regions of our synthetic observations that are far from the central source.

Positive bias affects polarization measurements because Stokes $Q$ and $U$ parameters can take either positive or negative values, but polarized intensity is always positive. This bias can be corrected using the methods outlined in \citet{Vaillancourt2006} and \citet{Hull2015}. We have not corrected for this bias in our polarization measurements; however, we neglect measurements at low signal-to-noise where the bias is significant, i.e., where $\bar{P} < 3\,\sigma_{P}$.

\begin{figure*}[hbt!]
\centering
\includegraphics[width=0.9\textwidth, clip, trim=2.5cm 0.0cm 0.0cm 0.0cm]{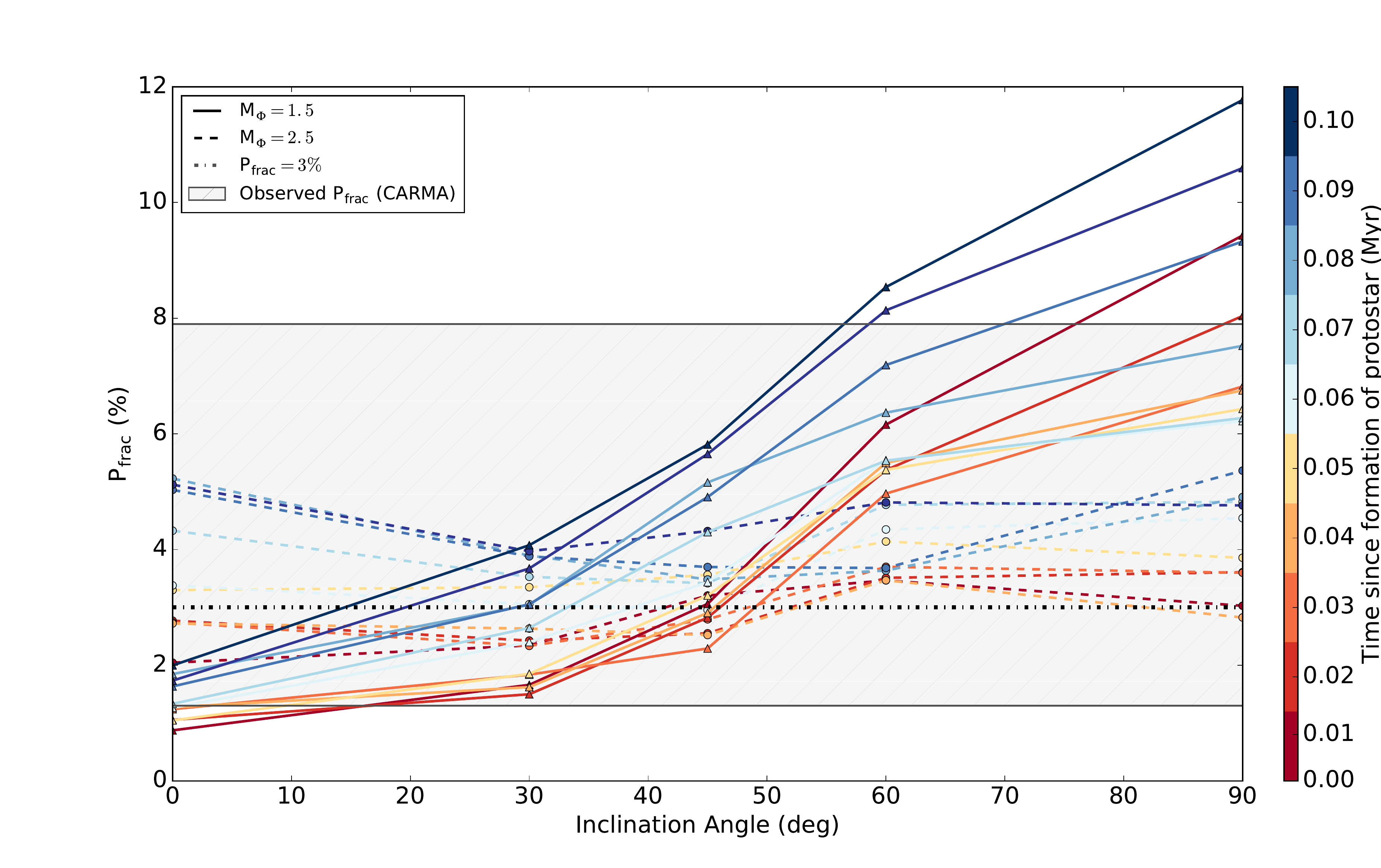}
\caption{\footnotesize
Polarization fraction as a function of viewing inclination angle, from 0\degree (face-on) to 90\degree (edge-on, viewed from the $y$-direction). The solid and dashed lines represent the two simulations, M$_\Phi=1.5$ and M$_\Phi=2.5$, respectively. The color scale indicates the age of the protostar, rather than the time passed since the beginning of the simulation. The hatched shaded area shows the range of polarization fractions observed by CARMA in \citet{Hull2014}. The thick, black dashed line at $\bar{P}_{\textrm{frac}} = 3$\% indicates the cutoff between low- and high-polarization sources.}
\vspace{0.2 in}
\label{fig:Pfracincl}
\end{figure*}

We compare the fractional polarization in the synthetic dust polarization maps across various inclination angles. In Figure~\ref{fig:Pfracincl}, we plot the polarization fraction as a function of viewing angle, for $0\degree$ (face-on), $30\degree, 45\degree, 60\degree$, and $90\degree$ (edge-on). The data include the synthetic sources in the Class 0 regime from both simulations. However, we ignore cases where the outflow is not distinguishable clearly by eye because the comparable observational study selected sources on the basis of clear bipolar outflows.

Figure~\ref{fig:Pfracincl} shows that the sources with a higher initial mass-to-flux ratio (weaker magnetic field) have lower fractional polarization as well as a smaller range of polarization fraction values over time. These less magnetized sources are also less affected by inclination angle and exhibit a smaller gradient in polarization fraction vs. viewing angle in all cases. In the highly magnetized case, the gradient of $\bar{P}_{\textrm{frac}}$ from low to high inclination angles is much steeper for all times.  This increase in polarization fraction as a function of inclination angle is consistent with the findings of \citet{Kataoka2012}, who performed synthetic observations of rotating, magnetized, non-turbulent protostellar cores.

\begin{figure}[hbt!]
\centering
\includegraphics[width=0.49\textwidth, clip, trim=0.4cm 0cm 0.4cm 0cm]{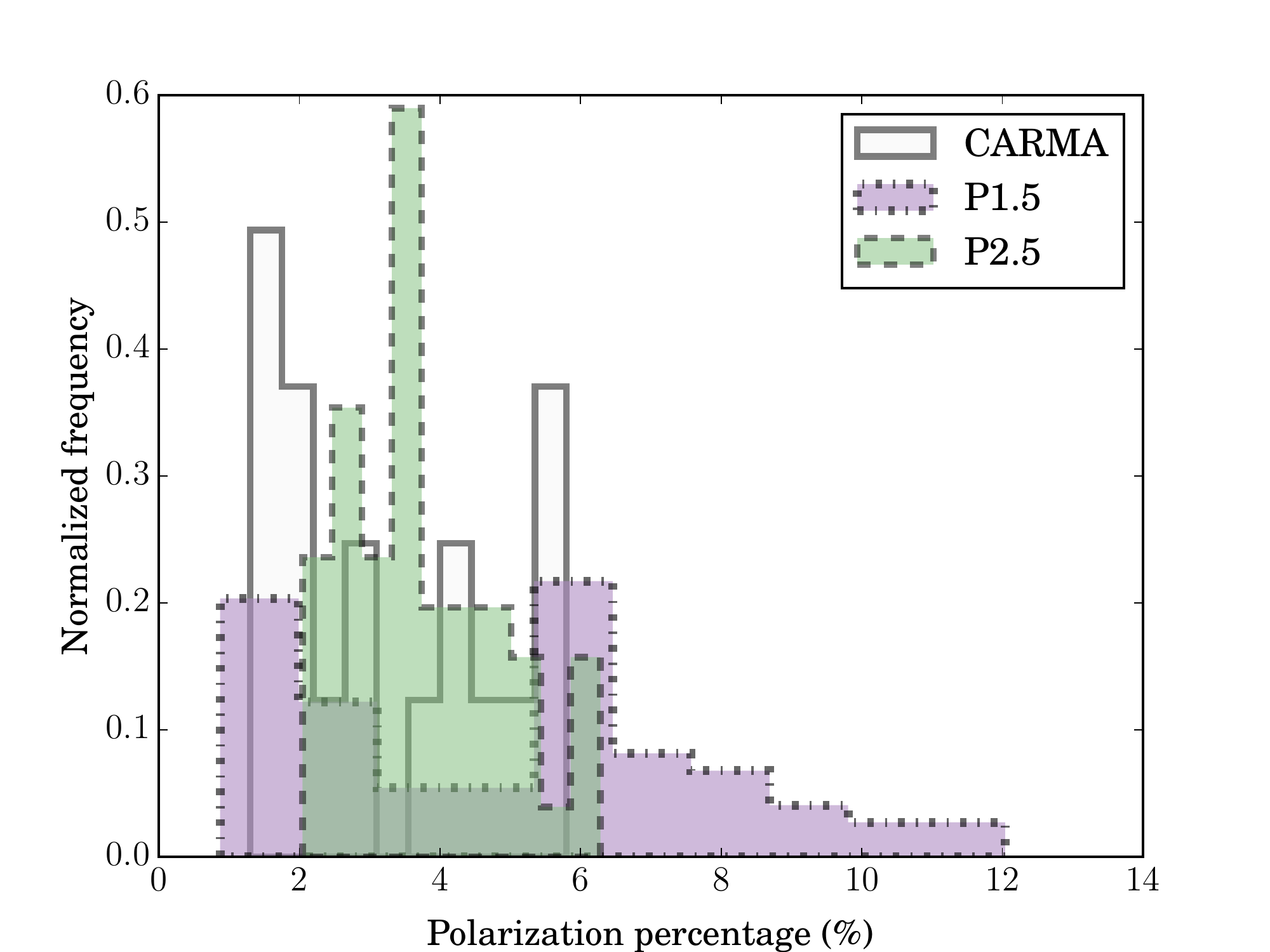}
\caption{\footnotesize
Histograms of the polarization fraction of the ``sources'' from the two simulations discussed in Section \ref{polfrac}, and from the CARMA TADPOL survey \citep[][Table 1]{Hull2014}.  The dot-dashed and the dashed histograms correspond to the various views of the strongly magnetized (S1.5) and weakly magnetized (S2.5) simulations, respectively.  The solid histogram comprises data from the TADPOL survey, and includes low-mass (Class 0, 0/I, or I) sources with polarization fraction detections above the 1.3\% CARMA detection limit.
}
\vspace{0.2 in} 
\label{fig:pfrac_hist}
\end{figure}

We analyze a sample of 126 views\footnote{A ``view'' is a rendering of the output (produced by \texttt{yt}) or a synthetic observation (produced by \texttt{DustPol}) at a given time step and at a given viewing angle.} from the two simulations that are equally distributed in time and are from a variety of viewing angles (i.e., the three orthogonal directions $x,$ $y,$ $z$ as well as views with inclination angles of 30$\degree$, 45$\degree$, and 60$\degree$).  There were 89 high-polarization views, 31 low-polarization views, and 6 non-detections. As in the CARMA survey, the cutoff between low and high polarization is defined as $\bar{P}_{\textrm{frac}}=3\%$, and non-detections are views with $\bar{P}_{\textrm{frac}} < 1.3\%$ (i.e., below the detection limit of the survey). Of the 89 high-polarization views, 31 lie above the maximum $\bar{P}_{\textrm{frac}} = 5.8\%$ detected by CARMA in a low-mass protostellar source.

The non-detections are all face-on views from S1.5.  A source with a perfectly vertical magnetic field would appear to have no magnetic flux when viewed along the field lines (i.e., when viewed from the face-on direction in our simulations). This is true at the outset of both simulations. However, in a highly magnetized source, the strong magnetic field better retains its vertical structure over time compared with a less magnetized source. This could explain why face-on views from S1.5 are not detected, but those from S2.5 are.

The 31 cases that lie above the maximum $\bar{P}_{\textrm{frac}}$ detected by CARMA (5.8\%) are all from views with inclinations that are close to edge-on, which suggests that sources observed to have high polarization fractions are likely to have ordered magnetic fields that lie close to the plane of the sky.

To illustrate the general trends in polarization fraction, in Figure \ref{fig:pfrac_hist} we have plotted histograms of the polarization fraction for the S1.5 views, the S2.5 views, and the data from the CARMA TADPOL survey. As we note in Section \ref{sec:method3}, it is important to remember that the fractions of high- and low-polarization views that we quote above will scale with the grain alignment efficiency (see the Appendix).  Assuming that 15\% is a reasonable value for the maximum grain-alignment efficiency, we see that the strongly magnetized simulation (S1.5) has far more high-polarization ``sources'' than either the weakly magnetized simulation (S2.5) or the CARMA data, suggesting that S1.5 may be more strongly polarized than typical protostellar sources such as those observed by \citet{Hull2014}.

We also investigate the time evolution of the polarization fraction and find that the polarization fraction tends to increase with time, particularly in the S2.5 case. This is illustrated in Figure \ref{fig:Pfracincl} by the color of the lines. The color bar on the right shows the age of the protostar in Myr. The oldest protostars (blue) have higher polarization fractions than the younger protostars (red). In contrast, observed Class I and Class II sources, which are expected to be older on average than Class 0 sources, tend to have lower polarization fractions, suggesting that either magnetic fields become less orderly over time or that magnetic fields weaken due to non-ideal MHD effects.

Many of the synthetic observations show the fractional ``polarization hole,'' a well known phenomenon where the polarization fraction drops at the dust emission peak. Examples that show the polarization hole include the central and bottom panels of Figure \ref{fig:Energy density 304560}. This appears in both high- (2.5$\arcsec$) and low-resolution (20$\arcsec$) observations \citep{1996ApJ...470..566D, Girart2006, 0004-637X-771-1-71} and simulations \citep{0004-637X-559-2-1005, Lazarian2007, Pelkonen2009}. Three possible reasons and explanations are discussed in \citet{Hull2014}. First, where the magnetic field is ordered along the line of sight due to turbulence or rotation, averaging along the line of sight reduces the overall polarization fraction. Second, the plane-of-sky magnetic field could have structure on scales <\,2.5$\arcsec$ that cannot resolved by CARMA; this plane-of-sky averaging would also reduce the polarization fraction. Finally, it is possible that poor alignment of grains at the center of the cores could arise due to less efficient grain alignment in regions of high extinction or collisions which knock grains out of alignment at high densities.

\begin{figure*}[hbt!]
\centering
\includegraphics[width=0.98\textwidth, clip, trim=0.2cm 0cm 0cm 0cm]{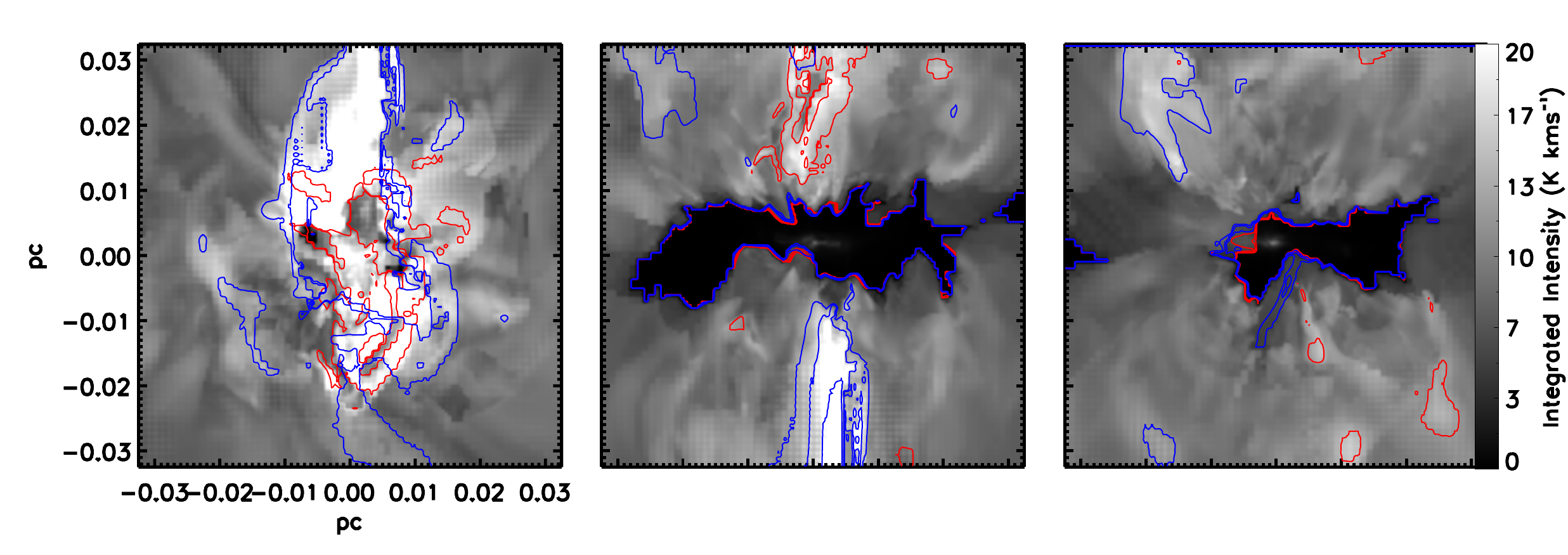}
\caption{\footnotesize
Synthetic integrated $^{12}$CO(2--1) intensity for $x-y$ (left), $y-z$ (middle) and $y-z$ (right) views at time 0.287 Myr (same perspectives and time as Figure 3). The contours denote the first moment of the $^{12}$CO(2--1) intensity ($\int I\,v\,dv/\int I\,dv$). The contours are [-2.5, -1.25, 1.25, 2.5] km s$^{-1}$, [-2.5, -1.25, -0.75, 0.75, 1.25, 2.5] km s$^{-1}$, and [-2.5, -1.25, 1.25, 2.5] km s$^{-1}$ for the left, middle and right panel, respectively. Receding gas is red and approaching gas is blue.
} 
\vspace{0.2 in}
\label{fig:cooutflow}
\end{figure*}

\subsection{Outflows}
\label{sec:outflows}

For each output, we calculate the outflow orientation and average magnetic field orientation viewed from different viewing angles and compare them. We ignore cases where the outflow is not clearly distinguishable by eye as these are unlikely to be identified in observations. We define the outflow angle using the average orientations of the primary bipolar lobes, and we define the magnetic field orientation as the unweighted circular average of the magnetic field orientations where $I > 3\,\sigma_{I}$ and $P > 3\,\sigma_{P}$. 

The  $^{12}$CO(2--1) emission line is a good tracer of warm outflow material, and consequently, it highlights the outflow morphology.  Figure \ref{fig:cooutflow} shows synthetic CO observations for three views at time 0.287 Myr, which are the same views depicted in Figure 3.  The first moment of the emission, which serves as a proxy for the outflow velocity, is overlaid.  The central part of the core is cold and dense and, thus, exhibits little CO(2--1) emission in the edge-on views. In the $x-y$ (face-on) view, the red- and blue-shifted gas overlaps such that the outflow direction is unclear, similar to the view in the top panel of Figure 3. Likewise, the CO emission in the other two views correlates well with the actual outflow direction determined from the raw gas distribution. These synthetic observations underscore that our outflow identification gives similar results to that performed using molecular line observations.

To check for a possible coupling between the outflow and the magnetic field, we compare the changes in the orientations of the outflow and the field over time, looking for any correspondence between their behaviors.  We define the change in orientation as the change of the magnetic field or outflow orientation with respect to the initial orientation. We presume that both the field and the outflow always move through an acute angle between one orientation and the next, i.e., neither the field nor the outflow rotates by over 90$\degree$ between subsequent time steps.

\begin{figure*}[hbt!]
\centering
\includegraphics[width=0.48\textwidth, clip, trim=1cm 0.7cm 2.4cm 0.9cm]{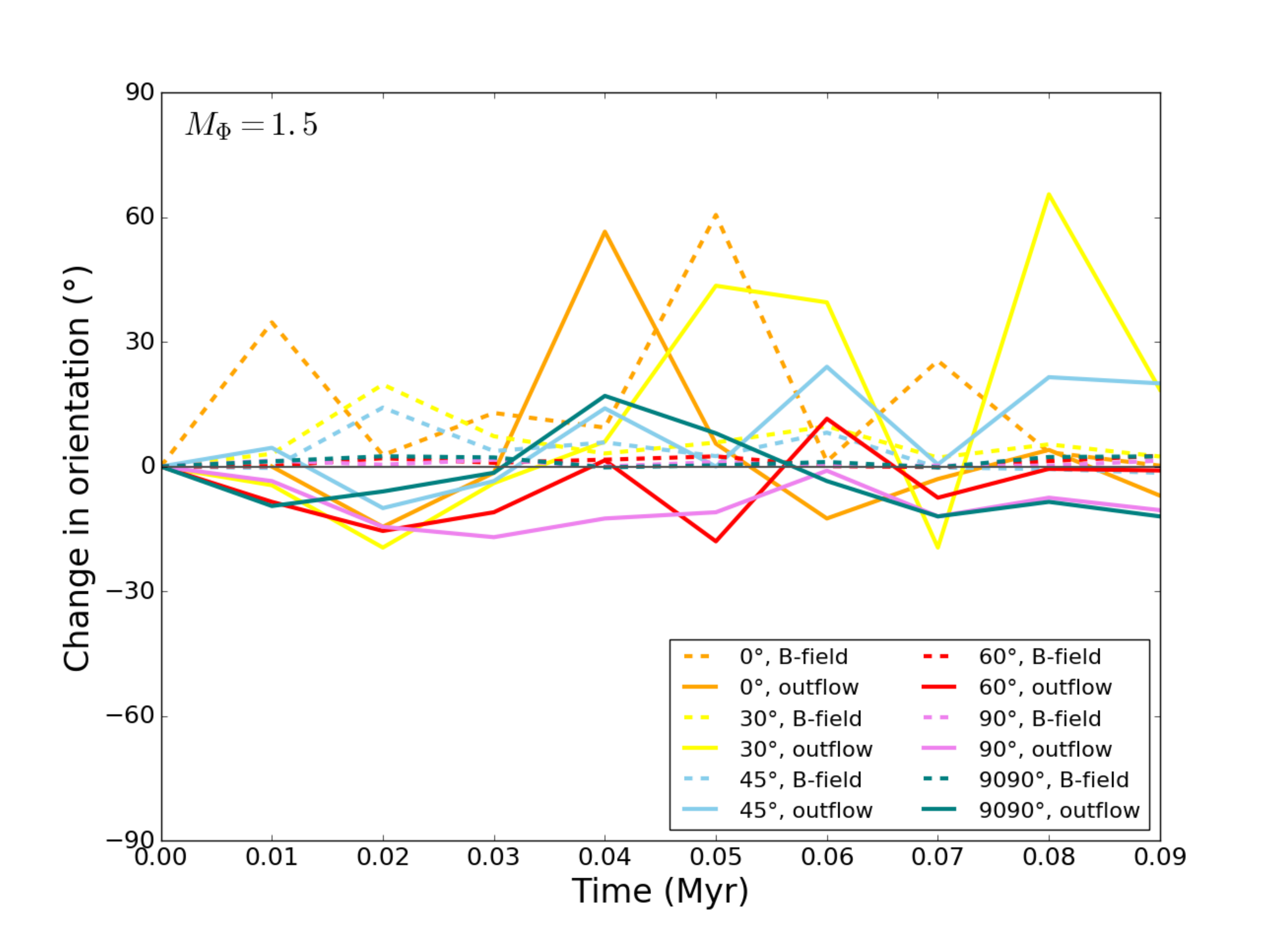}
\includegraphics[width=0.48\textwidth, clip, trim=1cm 0.6cm 2.4cm 0.9cm]{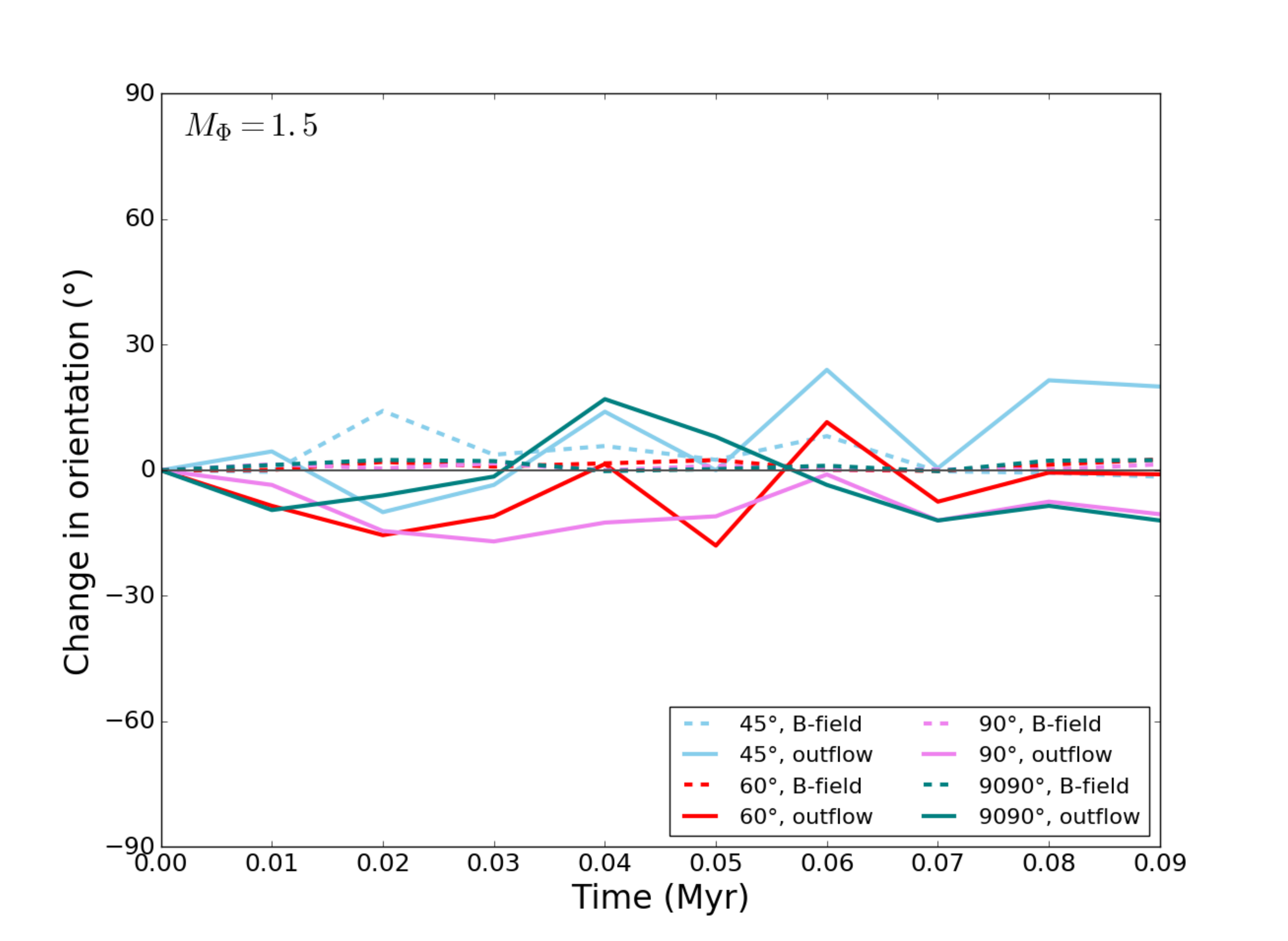}
\includegraphics[width=0.49\textwidth, clip, trim=1cm 0.7cm 2.4cm 0.9cm]{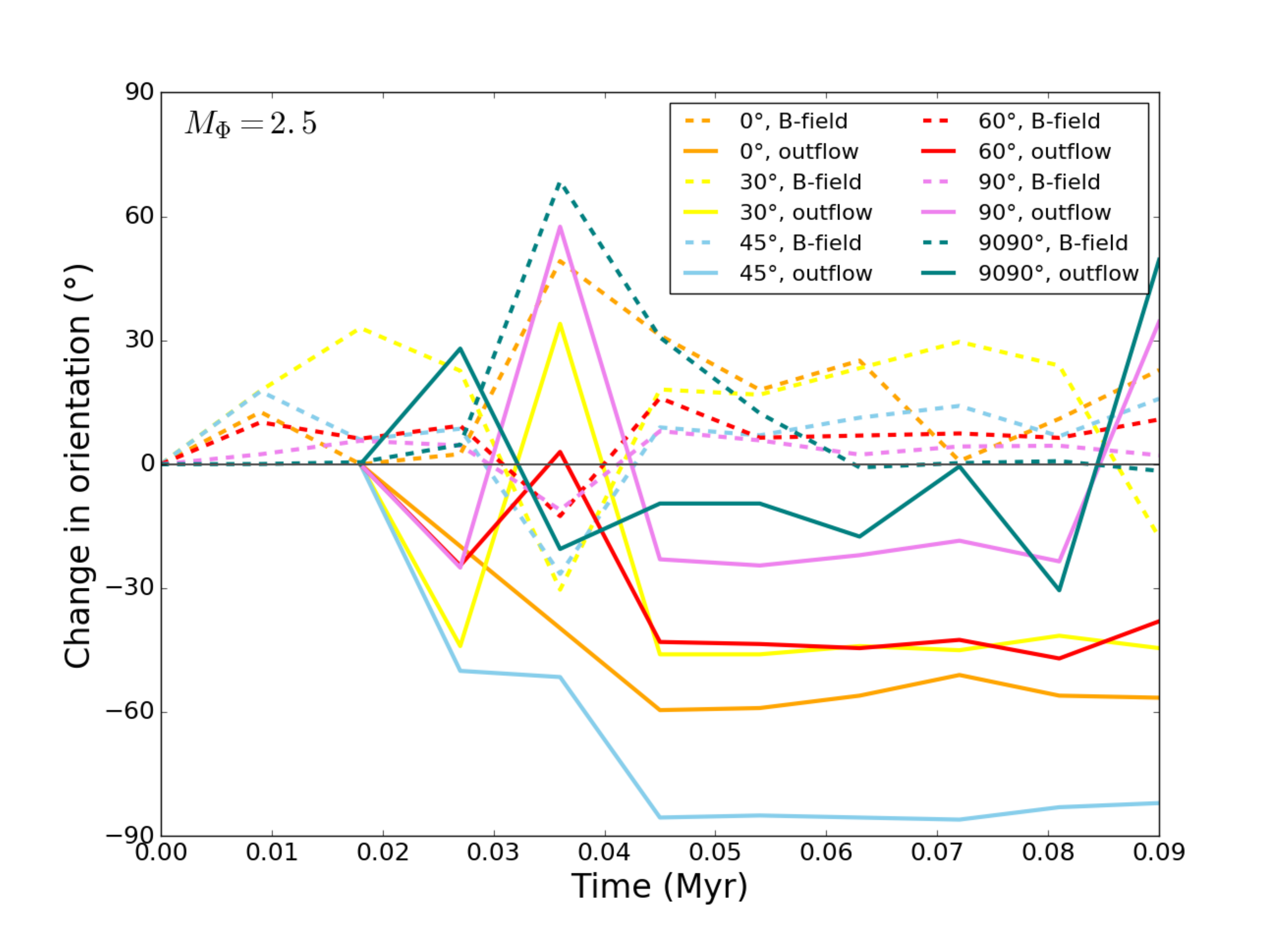}
\includegraphics[width=0.49\textwidth, clip, trim=1cm 0.6cm 2.4cm 0.9cm]{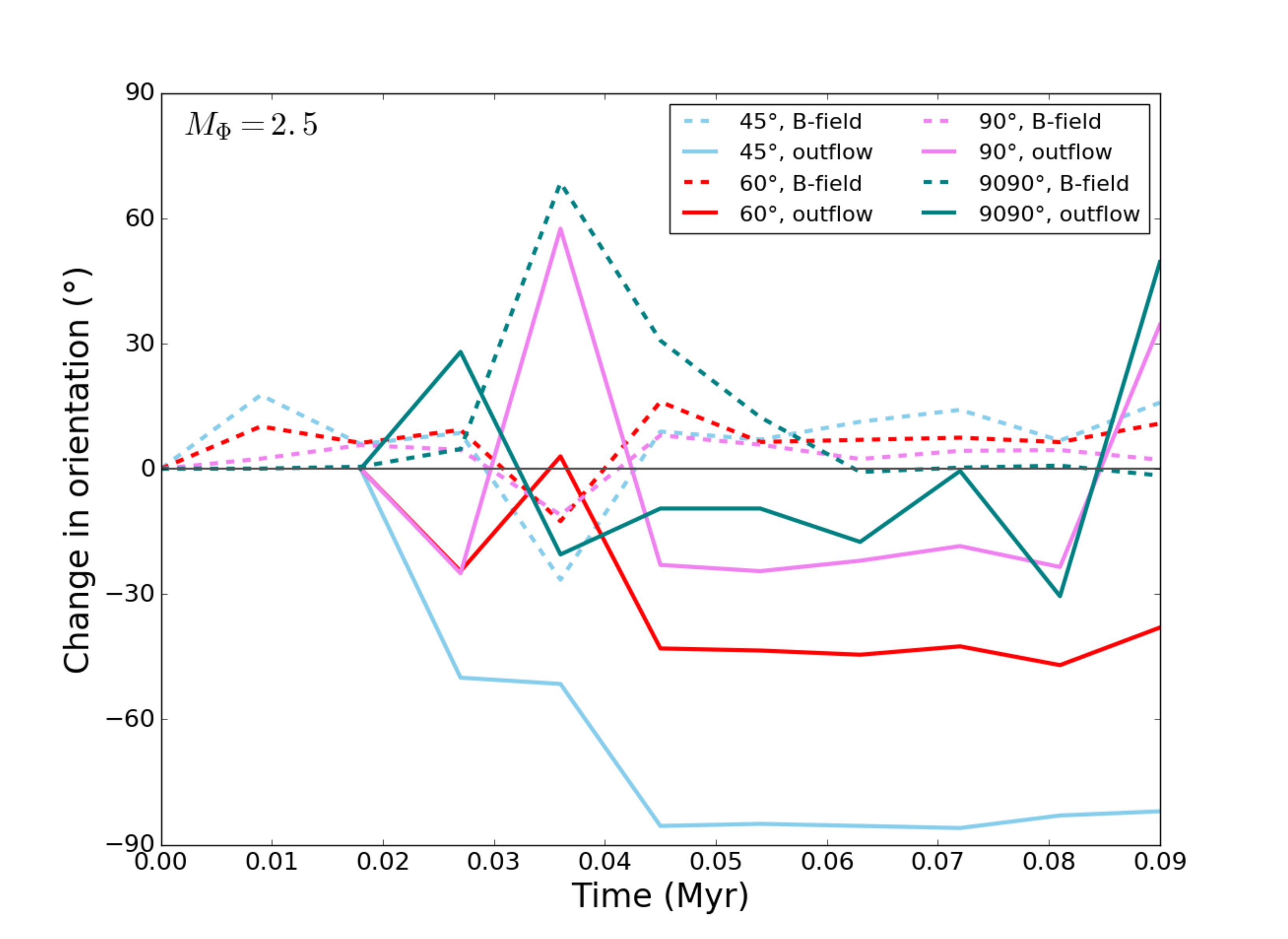}
\caption{\footnotesize
\textit{Top:} change in outflow orientation (solid lines) and change in magnetic field orientation (dashed lines) as a function of time for the highly magnetized core ($M_\Phi=1.5$).  \textit{Bottom:} same for the less magnetized core ($M_\Phi=2.5$).  The orientation angles are measured with respect to the initial direction of the magnetic field or outflow, showing the angular dispersion of the field or outflow over a 0.09\,Myr time period. \textit{Left:} outflows and magnetic fields viewed from 6 inclination angles, including face-on. \textit{Right:} outflows and magnetic fields viewed from 45--90$\degree$, i.e., where the magnetic field lies (nearly) in the plane of the sky.} 
\vspace{0.2 in}
\label{fig:Panglestime}
\end{figure*}

Figure \ref{fig:Panglestime} shows changes in magnetic field orientation and outflow orientation as a function of time (with a time step of 0.1\,Myr) for both S1.5 and S2.5. The evolutionary sequence shows visually that the outflow orientation changes significantly over the first 0.1\,Myr of the protostar's lifetime. A video of the evolution of the outflow (traced by energy density) in the S2.5 simulation can be found online.\footnote{ Video of the time evolution of the S2.5 simulation: \\\url{https://vimeo.com/134145942}).}

In the highly magnetized simulation, S1.5, the changes in the angles of both the outflow and the magnetic field are smaller than in S2.5. When we remove the nearly face-on data ($0 < \theta < 30\degree$) we see clearly that the magnetic field in the plane of the sky does not change orientation significantly over the Class 0 lifetime in S1.5, remaining within $30\degree$ of its initial direction (see the top right panel of Figure \ref{fig:Panglestime}). 
In contrast, the angle of the outflow in S2.5 \textit{does} change angle significantly over the first 0.05\,Myr, even when the face-on data are removed (see Figure \ref{fig:Panglestime}, bottom-right panel).

We should note that the system in S2.5 is a binary at early times. The merger of the primary and secondary at 0.38\,Myr produces a large change in the primary outflow direction due to the addition of the two protostellar angular momenta \citep[e.g.,][]{Fielding15}.
In the latter 0.05 Myr, the change in orientation plateaus in all planes, indicating that the outflow settles in one direction (between 0.05 and 0.08\,Myr). This suggests that binarity may also impact the degree of alignment between the field and outflow. Wide binary outflows are often misaligned with each other \citep{Lee2016, Offner2016}, which suggests that misaligned outflows and magnetic fields are probably common in binary or multiple systems. This follows logically since both of the outflows can't be misaligned with each other while also both being aligned with the magnetic field. A simulation with the same magnetic field strength as S2.5 but no initial turbulence shows much less variation in the outflow orientation (Offner et al. in prep.). This is consistent with prior laminar studies \citep{Kataoka2012}. 

We propose the following explanation for the behavior of the two simulations: in the highly magnetized case (S1.5), since the magnetic field is stronger, angular momentum is more efficiently removed from the turbulent gas, and consequently the net angular momentum of the accreting gas doesn't change much with time.  In contrast, in the less magnetized core (S2.5) the field and outflows actively change from all perspectives, at least for the first half of the time examined. While binarity may play a role in the misalignment (see above), this more lively behavior could also be explained by a weaker magnetic field, which would be less efficient at removing angular momentum (see next section). 

One caveat is that this study considers only two simulations. It is possible that the behavior of the outflow is less strongly coupled to that of the magnetic field and that the examples here coincidentally exhibit coupled behavior between magnetic fields and outflows. 

\subsection{Alignment of Fields and Outflows}

We revisit the question of misalignment between outflows and magnetic fields observed in Class 0 sources by \citet{Hull2014}. 
We plot cumulative distribution functions (CDFs) of the projected angle between the average magnetic field and outflow orientations and determine the degree of alignment statistically, by viewing each system from the positive $x$-, $y$- and $z$-directions. The measurement of average magnetic field and average outflow orientation is described in \S\ref{sec:outflows}; we define the difference between the magnetic field and outflow direction as the projected acute angle between them.

Following \citet{Hull2013, Hull2014}, we obtain the expected distributions of projected alignments using Monte Carlo models. These models simulate pairs of vectors in three dimensions that are tightly aligned ($0-20\degree$), somewhat aligned ($0-45\degree$), preferentially perpendicular ($70-90\degree$), and randomly aligned. The simulation then projects the vectors onto the plane of the sky and measures their angular differences in two dimensions.  

\begin{figure*}[hbt!]
\centering
\includegraphics[width=0.48\textwidth]{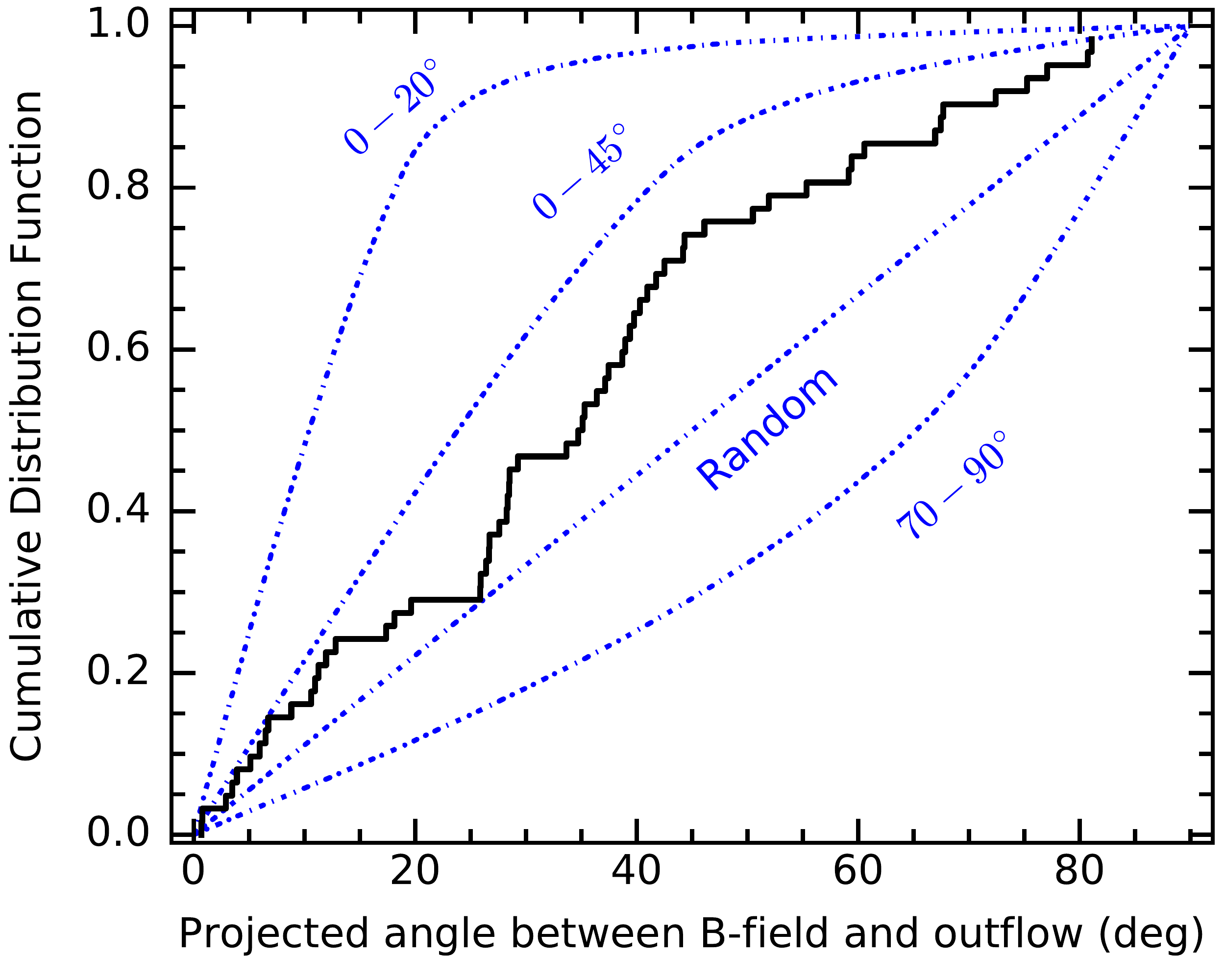}
\includegraphics[width=0.49\textwidth, clip, trim=0cm 0cm 0cm 0cm]{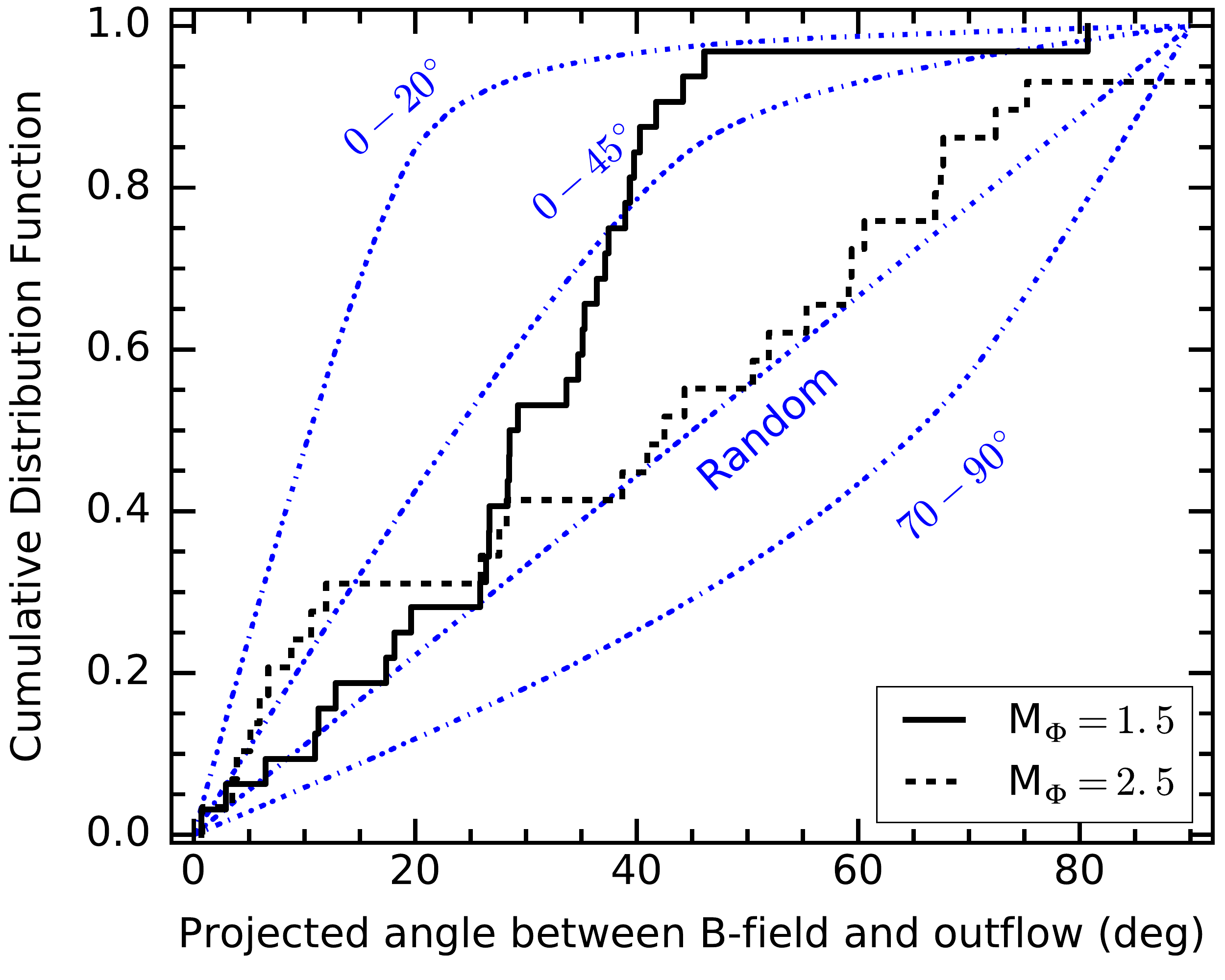}
\caption{\footnotesize
\textit{Left:} the stepped, solid black line shows the combined CDF of the (projected) angle between the outflow and magnetic field for all the synthetic sources that have clear outflows in both simulations. \textit{Right:} the CDFs for two simulations plotted separately. Angles from S1.5 are shown by the dotted black line (stepped) and for S2.5 by the dashed black line (stepped). Blue, dot-dashed lines indicate the CDFs of the projected outflow orientation differences generated by 3D Monte Carlo Simulations for tightly aligned ($0-20\degree$), somewhat aligned ($0-45\degree$), preferentially misaligned ($70-90\degree$), or randomly aligned (straight diagonal from bottom left to top right).} 
\vspace{0.2 in} 
\label{CDFall}
\end{figure*}

Figure~\ref{CDFall} shows the distributions for each type of alignment with blue dotted lines. The left panel of Figure~\ref{CDFall} shows the CDF for all the data including both of the simulations. Tight alignment and preferential misalignment can unequivocally be ruled out ($p<10^{-10}$). Both the somewhat aligned ($0-45\degree$) and randomly alignment distributions are also very statistically unlikely ($p=0.0004$ and $p=0.0012$, respectively). 

As the combined CDF does not conclusively show clear alignment or misalignment between magnetic fields and outflows, we consider the distributions for S1.5 and S2.5 separately. These distributions are shown in the right panel of Figure~\ref{CDFall}. In S1.5 (the highly magnetized core), there is more alignment between the magnetic field and the outflow than in S2.5 (the less magnetized core). The CDF for S1.5 is somewhat aligned ($0-45\degree$, with $p\approx0.01$); however, the CDF for S2.5 is best described by a random alignment ($p\approx0.3$); all other alignments can be ruled out. 

In ideal MHD, a more magnetized core with a greater magnetic field strength could be expected to remove angular momentum more efficiently, and therefore to show more alignment with the outflow. This explanation has some tentative support in the literature. \citet{Joos2013} performed MHD simulations of turbulent cores with misaligned magnetic fields and angular momentum vectors and found that stronger magnetic fields produced more collimated and higher-velocity outflows. However, they did not analyze the degree of alignment between the outflow and the magnetic field direction. \citet{Machida2006} and \citet{Kataoka2012} performed MHD studies of non-turbulent cores with different degrees of alignment between the rotation axis and the magnetic field direction.  As expected, they found that magnetic breaking was more efficient in the strong-field case, but that disks formed with their major axes perpendicular to the mean field even when the angular momentum vector was misaligned. This might suggest that outflows would also be parallel to the field (an outflow is produced in only one of their models), which contradicts the observations of \citet{Hull2013}.  \citet{Machida2013} also performed non-turbulent MHD simulations of protostellar cores and found that magnetic breaking suppressed outflow generation when the magnetic field was strong. However, since they did not include turbulence, the outflows are always aligned with the field direction. In our results, turbulence appears to be the key ingredient to produce misalignment between the outflow and field. However, exactly how this result depends on the mass-to-flux ratio requires a parameter study with a broader range of turbulence and field strengths.

Finally, we note that using only views from three directions has its limitations: first, this produces a statistically small number of ``sources,'' and second, intermediate views between edge-on and face-on are excluded. A sample with a larger number of isotropically distributed views would be ideal; however, analyzing views from isotropic points on a sphere is beyond the scope of this study.

\vspace{0.1in}
\section{Conclusions}
\label{sec:conclusions}

We have analyzed two ideal MHD simulations of protostellar cores with differing mass-to-flux ratios, and we have performed radiative transfer to model thermal dust emission. Having studied the qualitative and statistical properties of both the visualizations of simulation data and the synthetic dust polarization maps, we come to the following conclusions:

\begin{enumerate}
\item Polarization fraction increases when the system is viewed close to edge-on (where the magnetic field is close to the plane of the sky in our simulations), which suggests that sources observed to have high polarization fractions are likely to have ordered magnetic fields that lie close to the plane of the sky.  This is consistent with the synthetic observations performed by \citet{Kataoka2012}.

 \item The fractional polarization measurements from the simulation of the less magnetized core lie within the same range as observational measurements. In contrast, the fractional polarization measurements in the highly magnetized core frequently exceed the observed limits. This suggests that our synthetic ``sources''---in particular those derived from the more highly magnetized simulation with a mass-to-flux ratios of M$_\Phi$=1.5---may be more strongly magnetized than typical Class 0 sources, including those studied by \citet{Hull2014}.  However, we note that conclusions regarding the polarization fraction depend on the maximum grain alignment efficiency specified in \texttt{DustPol}.
 
\item Over the $\sim$\,0.1\,Myr Class 0 lifetime, the outflows move through large angles on relatively short ($\sim$\,0.01 Myr) timescales. In the less magnetized case, S2.5, the outflow and magnetic field orientations appear uncorrelated, suggesting that the direction of the angular momentum is not set by the magnetic field. However, there is more correlation between the outflow and magnetic field orientations in S1.5.

\item We construct CDFs of the projected angle between outflow angle and magnetic field orientations and find that S2.5 exhibits random alignment. This result is consistent with the observational survey by \citet{Hull2014}. S1.5 shows a higher degree of alignment, however, which is best described by models that are ``somewhat aligned'' between $0-45\degree$.  The better alignment of fields and outflows in the more magnetized simulation S1.5 is consistent with previous studies \citep[e.g.,][]{Matsumoto2006}; however, exactly how this result depends on the mass-to-flux ratio requires a parameter study with a broader range of turbulence and field strengths.
    
\end{enumerate}

Future work will improve upon this study specifically by exploring simulations with higher mass-to-flux ratios (i.e., less magnetized cores), different initial turbulent seeds, and a more diverse set of viewing angles.

Looking to the future of (sub)millimeter polarization observations: recent (as-of-yet unpublished) polarization data from ALMA have such high sensitivity that we are now able to probe the orientation of the magnetic field in forming stars on much smaller scales than were previously accessible to CARMA and the Submillimeter Array (SMA).  These new observations are enabling not only studies of polarization at the scale of protoplanetary disks (where the polarization may be due to a combination of thermal dust emission and scattering: see \citealt{Kataoka2015, Kataoka2016, Kataoka2016b, Yang2016a, Yang2016b}) but are also producing magnetic field maps with far more independent polarization detections across each individual source.  These latter maps can be used to study the interplay between magnetic fields and turbulence in low-mass star-forming cores; because of limited image sensitivities, previous interferometric studies have mainly focused on high-mass star-forming regions \citep[e.g.,][]{Houde2016}.  Maps with many independent detections can also be used to study the importance of magnetic fields in shaping structure in a gravitationally-bound star-forming environment using methods like the ``histogram of relative orientation,'' which studies the relative morphology of the magnetic field and the gas/dust \citep{Soler2013, PlanckXXXII}.  The methods used in this study to model and analyze dust emission from simulations allow us to put constraints on the role of magnetic fields in star formation and will be useful for interpreting the wealth of impending high-resolution, high-sensitivity observations by ALMA and other planned (sub)millimeter telescopes.

\acknowledgements

The authors thank the anonymous referee for the thorough comments, which improved the paper substantially.
The authors thank Marco Padovani for his help in fielding initial questions about \texttt{DustPol},  Peter K. G. Williams for his indispensable help with software installation, and Neosha Narayanan for developing the {\it yt} energy-density volume-rendering script.
J.L. acknowledges the Submillimeter Array for funding her participation in the AAS winter conference and funding observing travel.
S.O. acknowledges support from NASA grant NNX15AT05G. 
 The simulations were carried out at Yale University High Performance Computing Center and the Massachusetts Green High Performance Computing Center. The data analysis, images and animations were made possible by {\it yt} \citep{Turk2011}. The National Radio Astronomy Observatory is a facility of the National Science Foundation operated under cooperative agreement by Associated Universities, Inc.  This research made use of APLpy, an open-source plotting package for Python hosted at \url{http://aplpy.github.com}.

\bibliographystyle{apj}
\bibliography{biblio}

\appendix

While 15\% is thought to be a reasonable value for the maximum grain alignment efficiency, we nonetheless explore the dependence of the maximum polarization fraction on the maximum grain alignment efficiency.  We do this for both the edge-on and face-on cases and find that the edge-on cases (where the magnetic field is close to the plane of the sky) have higher polarization fractions for a given grain alignment efficiency.

\begin{figure}[hbt!]
\centering
\includegraphics[width=0.65\textwidth, clip, trim=0cm 0cm 0cm 0cm]{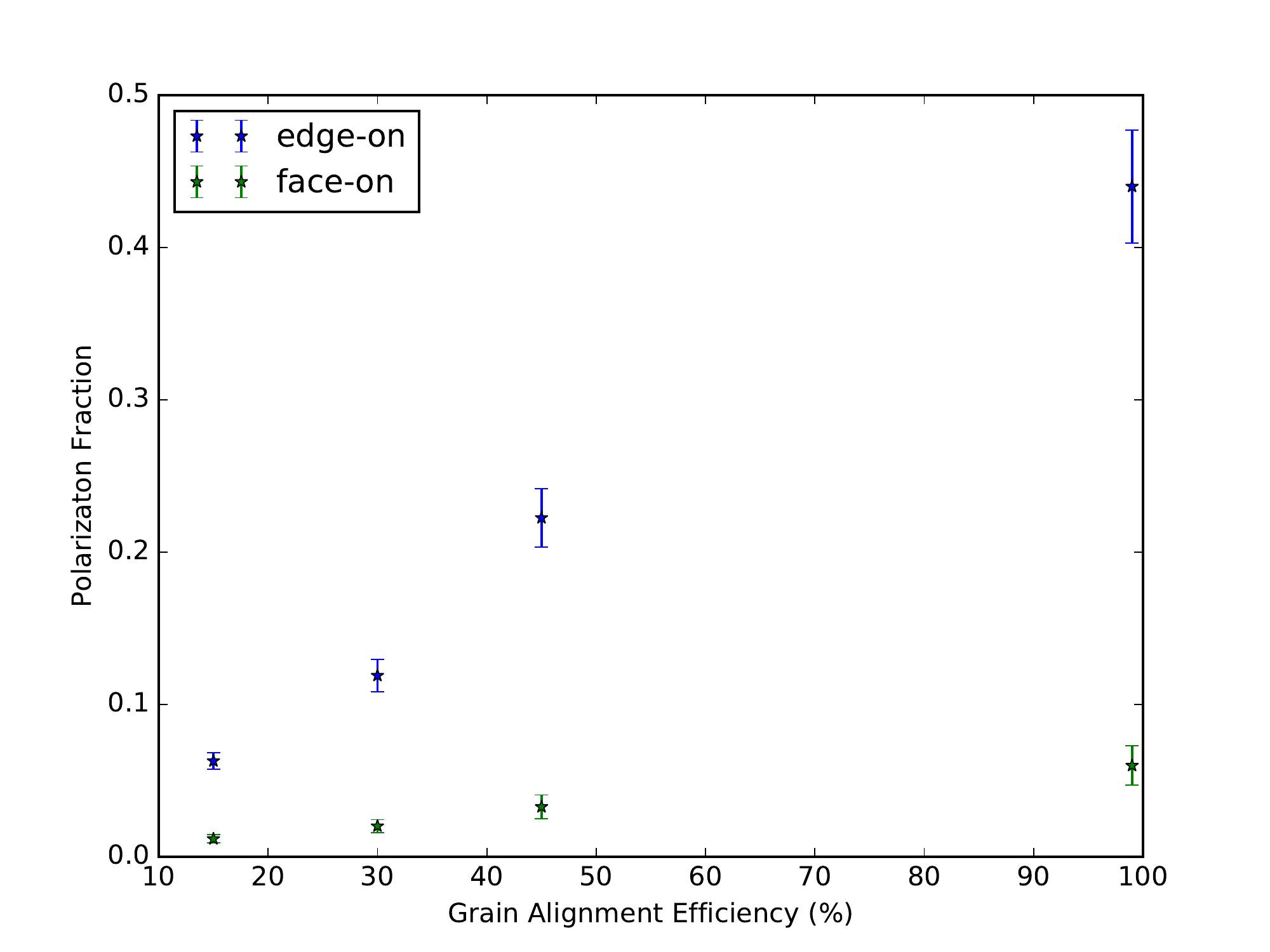}
\caption{\footnotesize
Maximum polarization fraction as a function of maximum grain alignment efficiency for both the edge-on and the face-on views of the S1.5 simulation at $t=0.353$\,Myr.} 
\vspace{0.2 in}
\label{fig:maxp}
\end{figure}

\end{document}